\documentclass[final,5p,times,twocolumn]{elsarticle}

\usepackage{amssymb}
\usepackage{amsmath}
\usepackage{bm}
\usepackage{comment}
\usepackage[subrefformat=parens]{subcaption}
\usepackage{xcolor}
\usepackage{hyperref}
\usepackage{slashed}
\hypersetup{
    colorlinks=true,       
    linkcolor=blue,          
    citecolor=blue,        
    urlcolor=blue,           
}

\journal{Physics Letters B}

\begin{document}

\begin{frontmatter}



\title{
  Anomalous enhancement of dilepton production
  as a precursor of color superconductivity
  \\ \vspace*{-12mm}\hspace{15.5cm} \small{\texttt{J-PARC-TH-0264}} \vspace*{10mm}
  }

\author[1]{Toru Nishimura\corref{cor1}}
\author[1,2]{Masakiyo Kitazawa}
\author[3]{Teiji Kunihiro}

\cortext[cor1]{Corresponding author}

\address[1]{Department of Physics, Osaka University, Toyonaka, Osaka, 560-0043 Japan}
\address[2]{J-PARC Branch, KEK Theory Center, Institute of Particle and Nuclear Studies, KEK, 319-1106 Japan}
\address[3]{Yukawa Institute for Theoretical Physics, Kyoto University, Kyoto, 606-8502 Japan}

\begin{abstract}
We compute the modification of the photon self-energy due to 
dynamical diquark fluctuations developed near the critical temperature of
the color superconductivity through 
the Aslamasov-Larkin, Maki-Thompson and density of states
terms, which are responsible for the paraconductivity in metals
at vanishing energy and momentum.
It is shown that the rate has a significant enhancement
at low invariant-mass region over a rather wide range of temperature
in the normal phase.
This enhancement is worth exploration in the relativistic heavy-ion collisions,
which may thereby reveal the significance of the diquark fluctuations 
in dense quark matter.
\end{abstract}


\end{frontmatter}



\section{Introduction}

Revealing the rich phase structure and thereby 
developing a condensed matter physics 
of Quantum Chromodynamics (QCD) in 
the high density region
is one of the main subjects in current nuclear physics~\cite{Fukushima:2010bq,Fukushima:2013rx}, and 
much endeavor has been being made 
both theoretically and experimentally.
In the high density region, for instance,
the first-order chiral transition line(s) with
the QCD critical 
point(s)  are  expected 
to exist on the basis of theoretical 
works~\cite{Asakawa:1989bq,Barducci:1989wi,Kitazawa:2002jop}, and
the experimental search for these phase transitions~\cite{Stephanov:1999zu}
is one of the  main purposes of
the beam-energy scan program in the relativistic heavy-ion collisions (HIC)
at RHIC, HADES and NA61/SHINE;
further 
studies to reveal the phase structure with higher statistics
will be pursued in future experiments 
planned at GSI-FAIR, NICA-MPD and J-PARC-HI~\cite{Galatyuk:2019lcf}.
Such studies on the Earth will also provide us with invaluable information
on the interior structure 
of compact stars~\cite{Kojo:2020ztt,Cierniak:2021knt,Kojo:2021wax}.

An interesting feature of the dense quark matter in yet higher density region
is the possible realization of the color superconductivity (CSC)
induced by
the condensation of diquark Cooper pairs~\cite{Alford:2007xm}.
Now that the future HIC experiments are designed so as to 
enable detailed analyses of the dense matter, 
it  would be intriguing to explore the possible existence of the CSC phases 
in these experiments.
The search for the CSC in the HIC, however, is quite a challenge
because the temperatures $T$ achieved in the HIC can become as high as
$100$~MeV at the highest baryon density~\cite{Ohnishi:2015fhj},
which may be  much higher than the critical temperature $T_c$ of 
the CSC, and hence an observation of 
the  CSC phases can be unlikely in the HIC.

Nevertheless, the  matter created in  the HIC may be within the critical 
region above $T_c$ where the diquark-pair fluctuations are
significant, and thus {\it precursory phenomena} of the
CSC~\cite{Kitazawa:2001ft,Kitazawa:2003cs,Kitazawa:2005vr}
do manifest themselves through appropriate observables by the HIC.
In this respect, it is suggestive that 
fluctuations of Cooper pairs (preformed pairs) of electrons in metals are
known to cause an anomalous enhancement of 
the electric conductivity above $T_c$ 
of the superconductivity (SC)~\cite{skocpol1975fluctuations,book_Larkin}.
Moreover, 
since the quark matter in the relevant density region is a strongly-coupled
system~\cite{Abuki:2001be,Kitazawa:2003cs}, 
the CSC can have a wider critical region where
the precursory phenomena of the CSC are pronounced.
In fact, 
it has been already shown~\cite{Kitazawa:2001ft,Kitazawa:2003cs,Voskresensky:2003wd,Kitazawa:2005vr,Kunihiro:2007bx,Kerbikov:2014ofa,Kerbikov:2020lqm}
that  the diquark fluctuations develop a well-defined collective mode,
which is the soft mode of the CSC, 
and its collectivity and the softening nature 
affect various observables
including the appearance of the ``pseudogap'' region~\cite{Kitazawa:2003cs}
in a rather wide range of temperature.

In the present Letter, 
we investigate possible enhancement of the production rate of virtual
photons due to the precursory diquark fluctuations,
which is to be  observed as the dilepton production rate (DPR) in the HIC.
A desirable feature of 
the electromagnetic probes, needless to say,  lies in the fact that
the interactions of the probes  with the medium are weak,
and  their properties are hardly modified from what they had when created, 
in contrast to hadronic signals.

Here we remark that the DPR 
{\em in} the CSC phases below $T_c$ is known to
show some unique behavior~\cite{Jaikumar:2001jq}.
However, such a behavior 
becomes weaker when $T$ goes higher and closer to $T_c$
because they are caused by the finite diquark gap.
On the other hand, the precursory phenomena to be 
investigated in the present Letter
are most enhanced at $T=T_c$, which is an attractive feature in the HIC.

The medium modification of the DPR or the 
virtual photons
is dictated by 
 that of the  photon self-energy~\cite{McLerran:1984ay,Weldon:1990iw,Kapusta:1991qp}.
The effects of the diquark fluctuations  on the photon self-energy
can be taken into account by
the Aslamasov-Larkin, Maki-Thompson and
density of states terms~\cite{Kitazawa:2005vr,Kunihiro:2007bx,Kerbikov:2020lqm}.
In the case of the metallic SC, 
these terms at the vanishing energy-momentum limit 
are known to explain an anomalous enhancement of the electric
conductivity above $T_c$~\cite{skocpol1975fluctuations,book_Larkin}.

In the present Letter, we calculate these terms composed of
diquark fluctuations near but above $T_c$ of the CSC at nonzero
energy and momentum.
We show that the Ward-Takahashi (WT) identity is satisfied
by summing up all of these terms.
From the imaginary part of these terms we calculate
a virtual photon emission from 
the diquark fluctuations that form a collective mode.
It is found that the virtual photons emitted from the collective mode
having a spectral support in the space-like region 
in turn have the spectral support in the time-like region.
Our numerical results show that the DPR
is significantly enhanced at low invariant-mass region $M\lesssim200$~MeV
above $T_c$ up to, say,  $T\simeq1.5T_c$, reflecting
the critical enhancement of the diquark fluctuations.
We argue that 
an experimental measurement of dileptons and exploration of the 
possible enhancement of the DPR in that far low-mass region  
in the HIC is quite worthwhile to do because  
it would  give an experimental 
evidence of strong diquark correlations
as a precurosr of 
the phase transition to CSC in dense quark matter.

\begin{figure}[tbp]
  \centering
  \includegraphics[width=0.4\textwidth]{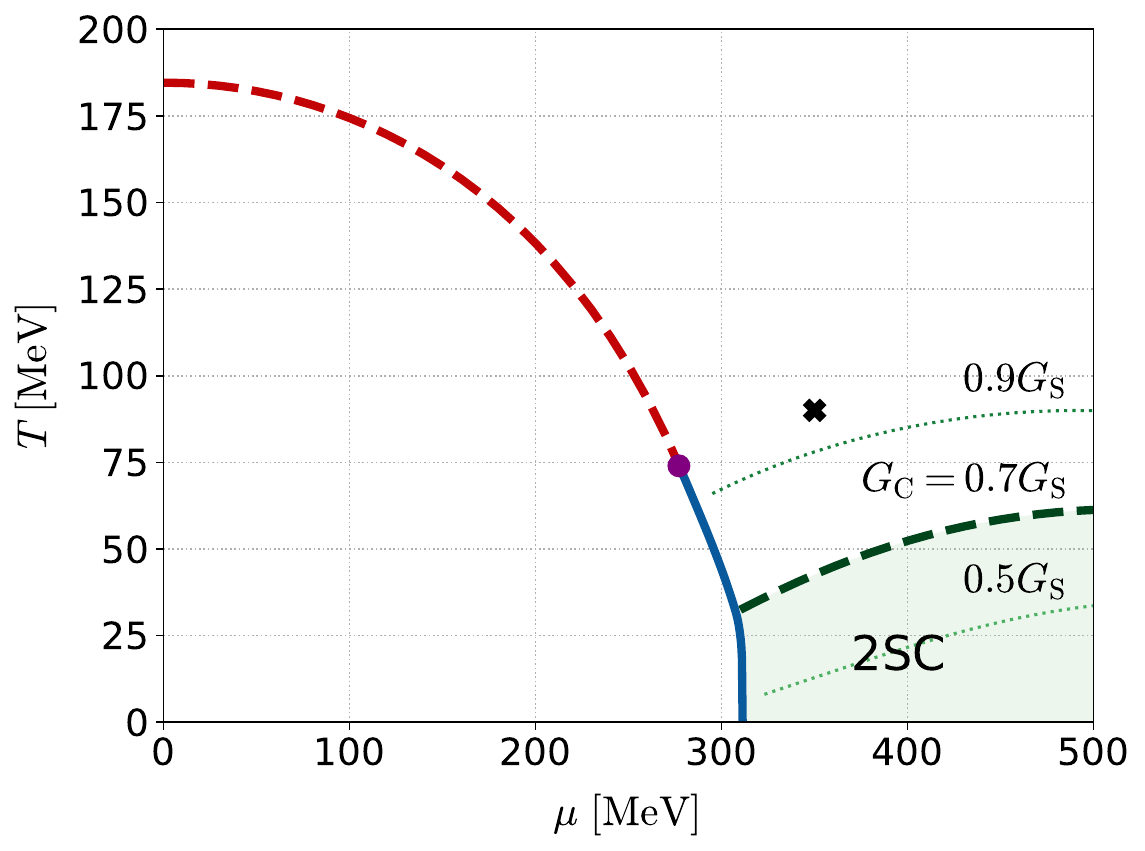}
\caption{
  Phase diagram obtained in the massless 2-flavor NJL model Eq.~(\ref{eq_lagrangian}).
  The bold lines show the transition lines at $G_{\rm C}=0.7G_S$.
  The solid and dashed lines represent the first- and second-order
  phase transitions.
  The $T_c$ of the 2SC phase at $G_{\rm C}=0.5G_S$ and $0.9G_S$ are also shown
  by the thin-dotted lines.
}
\label{fig_phase}
\end{figure}

\section{Model and phase diagram}
\label{sec:model}

In this Letter, we consider the diquark fluctuations above
$T_c$ of the 2-flavor superconductor (2SC), which is one of the
CSC phases that realizes at relatively low density~\cite{Alford:2007xm}.
We employ the 2-flavor NJL model~\cite{Hatsuda:1994pi,Buballa:2003qv} as
 an effective model of QCD to describe the phase transition to 2SC;
\begin{align}
  \mathcal{L} &= \bar{\psi} i 
  \slashed{\partial}
  \psi + \mathcal{L}_{\rm S} + \mathcal{L}_{\rm C},
  \label{eq_lagrangian}
  \\
  \mathcal{L}_{\rm S} &= \ G_{\rm S} [(\bar{\psi} \psi)^2 + (\bar{\psi} i \gamma_5 \vec{\tau} \psi)^2],
  \label{eq_LS}
  \\
  \mathcal{L}_{\rm C} &= \ G_{\rm C} (\bar{\psi} i \gamma_5 \tau_2 \lambda_A \psi^C)(\bar{\psi}^C i \gamma_5 \tau_2 \lambda_A \psi),
  \label{eq_LC}
\end{align}
where $\mathcal{L}_{\rm S}$ and $\mathcal{L}_{\rm C}$ represent the quark-antiquark 
and  quark-quark interactions, respectively, and
$\psi^C (x) = i \gamma_2 \gamma_0 \bar{\psi}^T (x)$.
$\tau_2$ and $\gamma_A$ $(A=2,5,7)$ are 
the antisymmetric components of the Pauli and Gell-mann matrices 
for the flavor $SU(2)_f$ and color $SU(3)_c$, respectively.
The scalar coupling constant $G_{\rm S}=5.01 \rm{GeV^{-2}}$ and the
three-momentum cutoff $\Lambda=650$~MeV are
determined so as to reproduce the pion decay constant $f_{\pi}=93 \rm{MeV}$ and 
the chiral condensate $\langle \bar{\psi} \psi \rangle = (-250\rm{MeV})^3$ in vacuum~\cite{Hatsuda:1994pi}.
The current quark mass is neglected for simplicity, while 
the diquark coupling $G_{\rm C}$ is treated as a free parameter.
We employ a common quark chemical potential $\mu$ for
up and down quarks since the effect of isospin breaking is not large
in the medium created in the HIC.

In Fig.~\ref{fig_phase}, we show the phase diagram in the $T$--$\mu$ plane
obtained in the mean-field approximation (MFA) with the mean fields
$\langle \bar{\psi} \psi \rangle$ and
$\langle \bar{\psi}^C \Gamma \psi \rangle~$
with $\Gamma = i \gamma_5 \tau_2 \lambda_A$.
The bold lines show the phase diagram at $G_{\rm C}=0.7G_{\rm S}$, where
the solid and dashed lines represent the first- and second-order
phase transitions, respectively.
The 2SC phase is realized in the dense region at relatively low temperatures.
In the figure, the phase boundary of the 2SC for $G_{\rm C}=0.5G_{\rm S}$ and $0.9G_{\rm S}$
is also shown by the thin-dotted lines.

In MFA, the phase transition to 2SC is of second order
as shown in Fig.~\ref{fig_phase}.
It is known that the transition becomes
first order due to the effect of gauge fields (gluons)
in asymptotically high density region~\cite{Matsuura:2003md, Giannakis:2004xt, Noronha:2006cz, Fejos:2019oxz}.
On the other hand, the fate of the transition at lower densities
has not been settled down to the best of the authors' knowledge.
In the present study we thus assume that the transition is second or
weak first order having the formation of the soft mode discussed below.

\section{The soft mode of 2SC}

\subsection{Propagator of diquark field}
\label{sec:propagator}

\begin{figure}[tbp]
\centering
\includegraphics[width=0.45\textwidth]{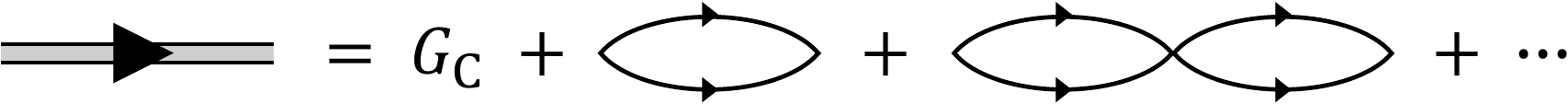}
\caption{
  Diagrammatic representation of the T-matrix Eq.~(\ref{eq:Xi})
  in the RPA.}
\label{fig_softmode}
\end{figure}

A characteristic feature of the second-order phase transition is that
the fluctuation amplitude of the order parameter diverges at $T=T_c$.
To see such a divergence at the $T_c$ of the 2SC,
let us consider the imaginary-time propagator of the diquark field
$\Delta(x)=\bar{\psi}^C(x) \Gamma \psi(x)$,
\begin {align}
  \mathcal{D}(k) 
  = -\int_0^{1/T} d\tau \int d^3\bm{x}
  \langle T_\tau \Delta^\dagger (x) \Delta(0)   \rangle e^{i \nu_l \tau} e^{-i \bm{k}\cdot\bm{x}}, 
  \label{eq:D}
\end {align}
where $k=(\bm{k}, i\nu_l)$ is the four momentum of the diquark field
with $\nu_l$ the Matsubara frequency for bosons, 
$\tau$ is the imaginary time, and $T_\tau$ denotes the imaginary-time
ordering.
In the random-phase approximation (RPA), Eq.~(\ref{eq:D}) is given by
$\mathcal{D}(k) = \mathcal{Q}(k)/(1+G_{\rm C}\mathcal{Q}(k))$ with
the one-loop quark-quark correlation function
\begin {align}
  \mathcal{Q} (k) = - 8 
  \int_p
  {\rm Tr} [\mathcal{G}_0 (k-p) \mathcal{G}_0(p)], 
  \label{eq:Q}
\end {align}
where $\mathcal{G}_0(p)
= 1/[(i\omega_m + \mu)\gamma_0 - \bm{p} \cdot \bm{\gamma}]$ is the free quark propagator with $p=(\bm{p},i\omega_m)$ and 
the Matsubara frequency for fermions $\omega_m$, 
${\rm Tr}$ denotes the trace over the Dirac indices,
and $\int_p = T\sum_m \int d^3 \bm{p}/(2\pi)^3$.
We also introduce the T-matrix to describe the diquark fluctuation 
\begin{align}
  \tilde\Xi(k)
  = \frac1{G_{\rm C}^{-1}+\mathcal{Q}(k)}
  = G_{\rm C} - G_{\rm C}\mathcal{D}(k)G_{\rm C},
  \label{eq:Xi}
\end{align}
which is diagrammatically represented in Fig.~\ref{fig_softmode}.

The retarded Green functions 
$D^R(\bm{k},\omega)$, $Q^R(\bm{k},\omega)$ and $\Xi^R(\bm{k},\omega)$
corresponding to Eqs.~(\ref{eq:D})--(\ref{eq:Xi}), respectively, are
obtained by the analytic continuation $i\nu_l\to\omega+i\eta$.
The imaginary part of $Q^R (\bm{k}, \omega)$ is calculated to be~\cite{Kitazawa:2005vr}
\begin{align}
  &{\rm Im} Q^R (\bm{k}, \omega) =
  -\frac{2T}{\pi k} [(\omega + 2\mu)^2 - k^2]
  \nonumber  \\
  &\qquad\qquad
  \times \bigg\{
  \log \frac{{\rm cosh}(\omega+k)/4T}{{\rm cosh}(\omega-k)/4T} 
  -\frac{\omega}{2T}
  \theta (k-|\omega+2\mu|)
  ~\bigg\} \ .
  \label{eq_ImQ}
\end{align}
Its real part is then constructed using the Kramers-Kronig relation 
\begin{align}
  {\rm Re} Q^R (\bm{k}, \omega) =\frac{1}{\pi} P \int^{2\Lambda-2\mu}_{-2\Lambda-2\mu} d\omega '
  \frac{{\rm Im} Q^R (\bm{k}, \omega)}{\omega '- \omega} \ ,
  \label{eq_ReQ}
\end{align}
where $P$ denotes the principal value~\cite{Kitazawa:2005vr}.

The retarded diquark propagator $D^R(\bm{k},\omega)$, and hence the T-matrix
$\Xi^R(\bm{k},\omega)$, has a pole at $\omega=|\bm{k}|=0$ at $T=T_c$;
$[D^R(\bm{0}, 0)]^{-1}_{T=T_c}=[\Xi^R(\bm{0}, 0)]^{-1}_{T=T_c}=0$.
This fact, known as the Thouless criterion~\cite{thouless1960perturbation},
is confirmed by comparing the denominator of $D^R(\bm{k},\omega)$
with the gap equation for the diquark field.
The criterion shows that the diquark field has a massless collective mode
at $T=T_c$.
Furthermore, the pole of this collective mode 
moves continuously toward the origin in the complex energy plane 
as $T$ is lowered to $T_c$, and hence the collective mode 
has a vanishing excitation energy toward $T_c$.
This collective mode is called the {\em soft mode}.
Because of the small excitation energy, they tend to be easily excited 
and affect various observables in the medium near $T_c$~\cite{Kitazawa:2003cs,Kitazawa:2005vr}.

Although we had recourse to the MFA and RPA,
the appearance of a soft mode is a generic feature of the second-order
phase transition~\cite{book_Larkin}, and 
even if the phase transition is 
of first order, 
the development of a collective mode with the softening nature 
prior to the critical point 
is still expected for weak first-order transitions.
Therefore, 
the emergence of the soft mode in the diquark channel and 
the following discussions on its effects on observables should
have a model-independent validity, at least qualitatively.

\begin{figure}[tbp]
\centering
\includegraphics[bb = 0 0 477 416, scale=0.45]{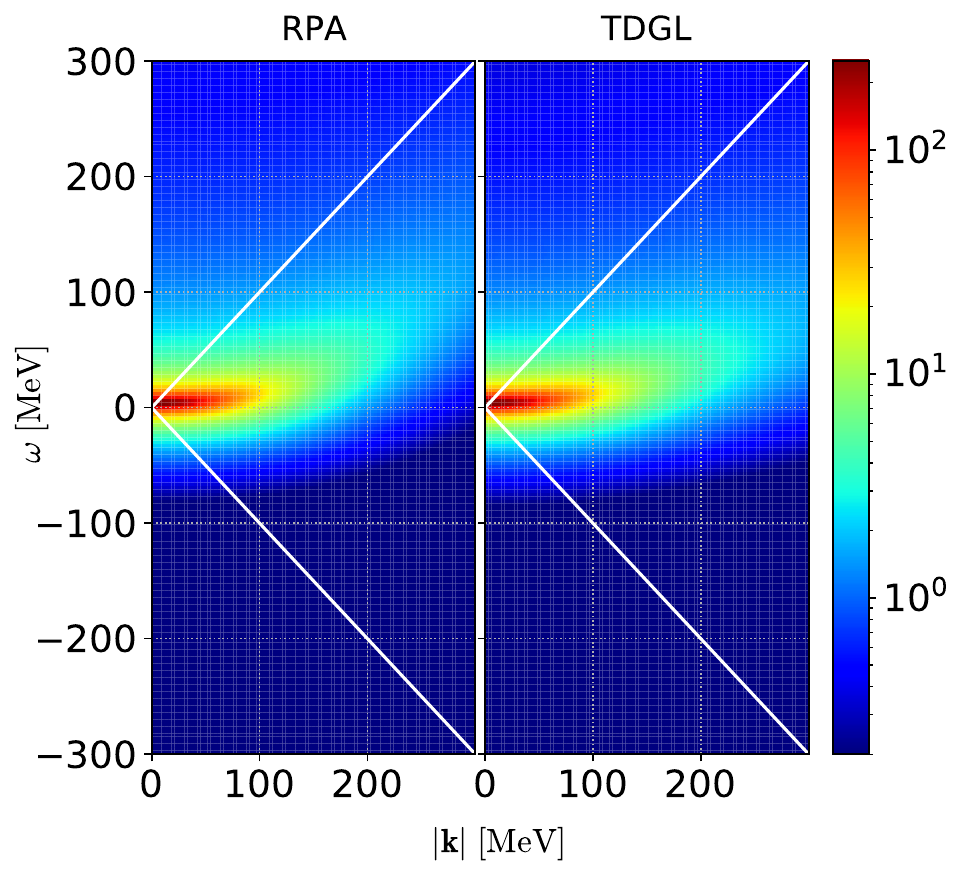}
\caption{
  Contour plot of the 
  dynamical structure factor $S(\bm{k},\omega)$
  at $T=1.05T_c$ for $\mu = 350$~MeV and $G_{\rm C}=0.7G_{\rm S}$.
  The solid lines show the light cone.
  The left panel is the result of RPA obtained from 
  Eqs.~(\ref{eq_ImQ}) and (\ref{eq_ReQ}),
  while the right panel is the result in the TDGL approximation
  Eq.~(\ref{eq_softmodeapprox}).
}
\label{fig_DSF}
\end{figure}

To detail  the properties of the soft mode,
it is convenient to introduce the dynamical structure factor 
$S(\bm{k},\omega)$ given by
\begin{align}
  S(\bm{k},\omega)
  = - \frac1\pi \frac1{1-e^{-\beta\omega}} {\rm Im}D^R (\bm{k}, \omega).
  \label{eq:S}
\end{align}
Figure~\ref{fig_DSF} shows a contour map of $S(\bm{k},\omega)$ at $T=1.05T_c$
for $\mu=350$~MeV and $G_{\rm C}=0.7G_{\rm S}$.
One sees that $S(\bm{k},\omega)$ has a clear spectral concentration with 
a peak 
around the origin in the $\omega$--$|\bm{k}|$ plane,
which implies a development of the collective mode having a
definite dispersion relation 
$\omega=\omega(|\bm{k}|)$ 
with a small width~\cite{Kitazawa:2001ft,Kitazawa:2005vr}.
We also note that the spectral concentration is confined in the space-like region,
$\omega(|\bm{k}|)<|\bm{k}|$.
This feature will be picked up again later when 
we discuss the DPR that has a spectral support in the time-like region.

\subsection{Time-dependent Ginzburg-Landau (TDGL) approximation}
\label{sec:TDGL}

Since the diquark fluctuations near $T_c$ have spectral concentration
in the low energy region as we have seen above,
we approximate the T-matrix $\Xi^R(\bm{k}, \omega)$
in the small $\omega$ region as
\begin {align}
  \Xi^R (\bm{k}, \omega) \simeq \frac{1}{c \omega + G_{\rm C}^{-1} + Q^R(\bm{k},0)} \ ,
  \label{eq_softmodeapprox}
\end {align}
with $c = \partial Q^R(\bm{0}, \omega)/ \partial \omega|_{\omega=0}$.
We refer Eq.~(\ref{eq_softmodeapprox}) to as the 
time-dependent Ginzburg-Landau (TDGL) approximation,
since Eq.~(\ref{eq_softmodeapprox}) corresponds to the linearlized
TDGL approximation for the T-matrix~\cite{Cyrot:1973}
without the expansion along $|\bm{k}|^2$.
In this study we do not expand $[\Xi^R(\bm{k}, \omega)]^{-1}$
with respect to $|\bm{k}|^2$
for a better description of the spectral strength
extending along $|\bm{k}|$ direction widely as in Fig.~\ref{fig_DSF}.
An explicit calculation shows that $c$ is a complex number,
while $G_{\rm C}^{-1} + Q^R(\bm{k},0)$ is real.

In the right panel of Fig.~\ref{fig_DSF}, we show $S(\bm{k},\omega)$
obtained by the TDGL approximation. 
By comparing the result with the left panel, one sees 
that the TDGL approximation Eq.~(\ref{eq_softmodeapprox}) reproduces
the result obtained by the RPA quite well in a
wide range in the $\omega$--$|\bm{k}|$ plane.

\begin{figure}[tbp]
\centering
\includegraphics[width=0.45\textwidth]{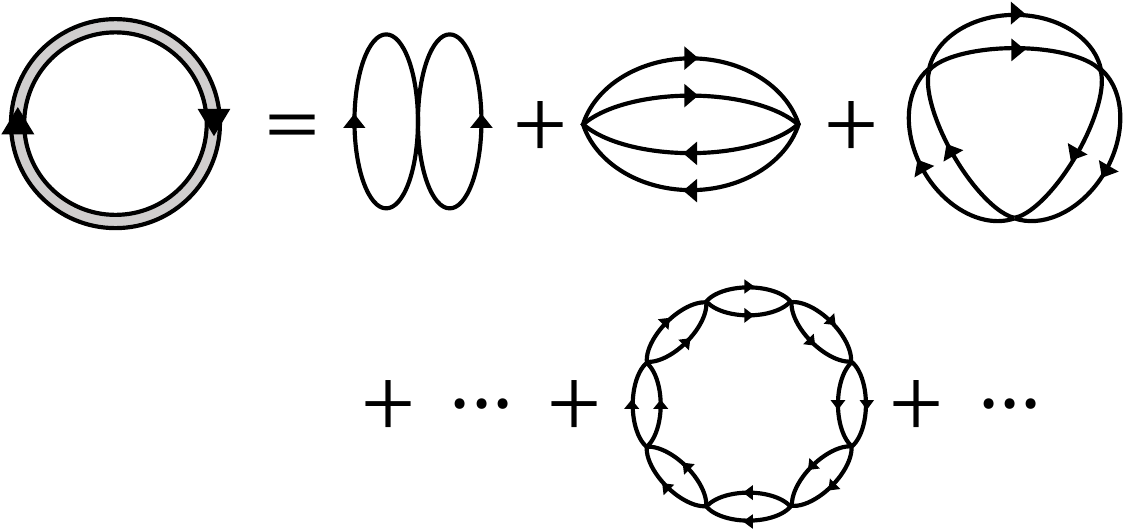}
\caption{Contribution of the diquark fluctuations to the thermodynamic potential.}
\label{fig_potential}
\end{figure}

\section{Photon self-energy and dilepton production rate}

The DPR
is given in terms of the retarded photon self-energy $\Pi^{R \mu\nu}(\bm{k}, \omega)$ as~\cite{McLerran:1984ay,Weldon:1990iw,Kapusta:1991qp},
\begin {align}
  \frac{d^4\Gamma(\bm{k}, \omega)}{d^4k} = -\frac{\alpha}{12\pi^4} 
  \frac{1}{\omega^2-|\bm{k}|^2} \frac{1}{e^{\beta\omega}-1} g_{\mu\nu}  
       {\rm Im} \Pi^{R \mu\nu} (\bm{k}, \omega) ,
       \label{eq_RATE_kom}
\end {align}
with the fine structure constant $\alpha$.

\subsection{Construction of the photon self-energy}

\begin{figure}[tbp]
    \begin{tabular}{cccc}
      \begin{minipage}[t]{0.45\hsize}
        \centering
        \includegraphics[bb = 0 0 917 424, keepaspectratio, scale=0.083]{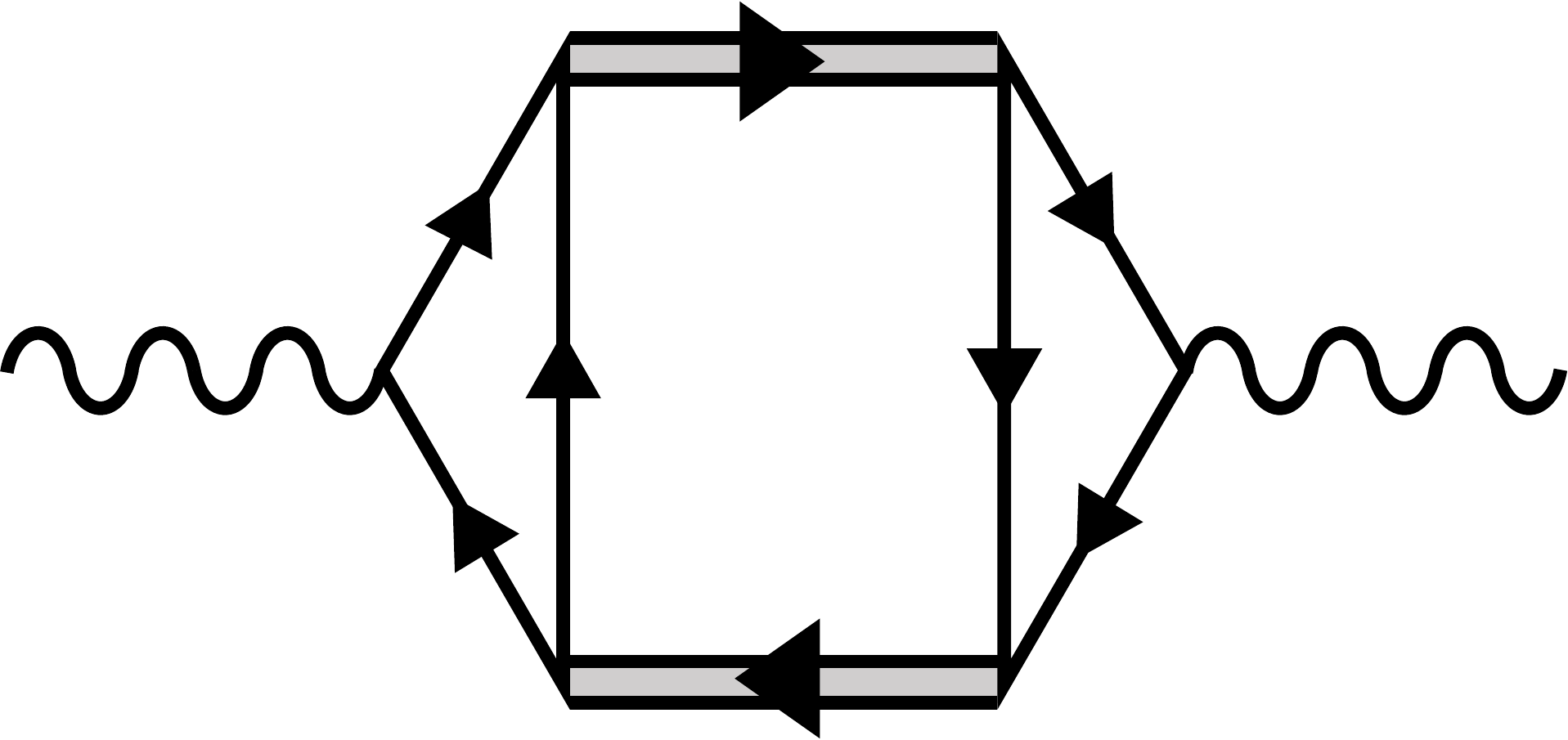}
        \subcaption{}
        \label{fig_AL}
      \end{minipage} &
      \begin{minipage}[t]{0.45\hsize}
        \centering
        \includegraphics[bb = 0 0 918 428, keepaspectratio, scale=0.083]{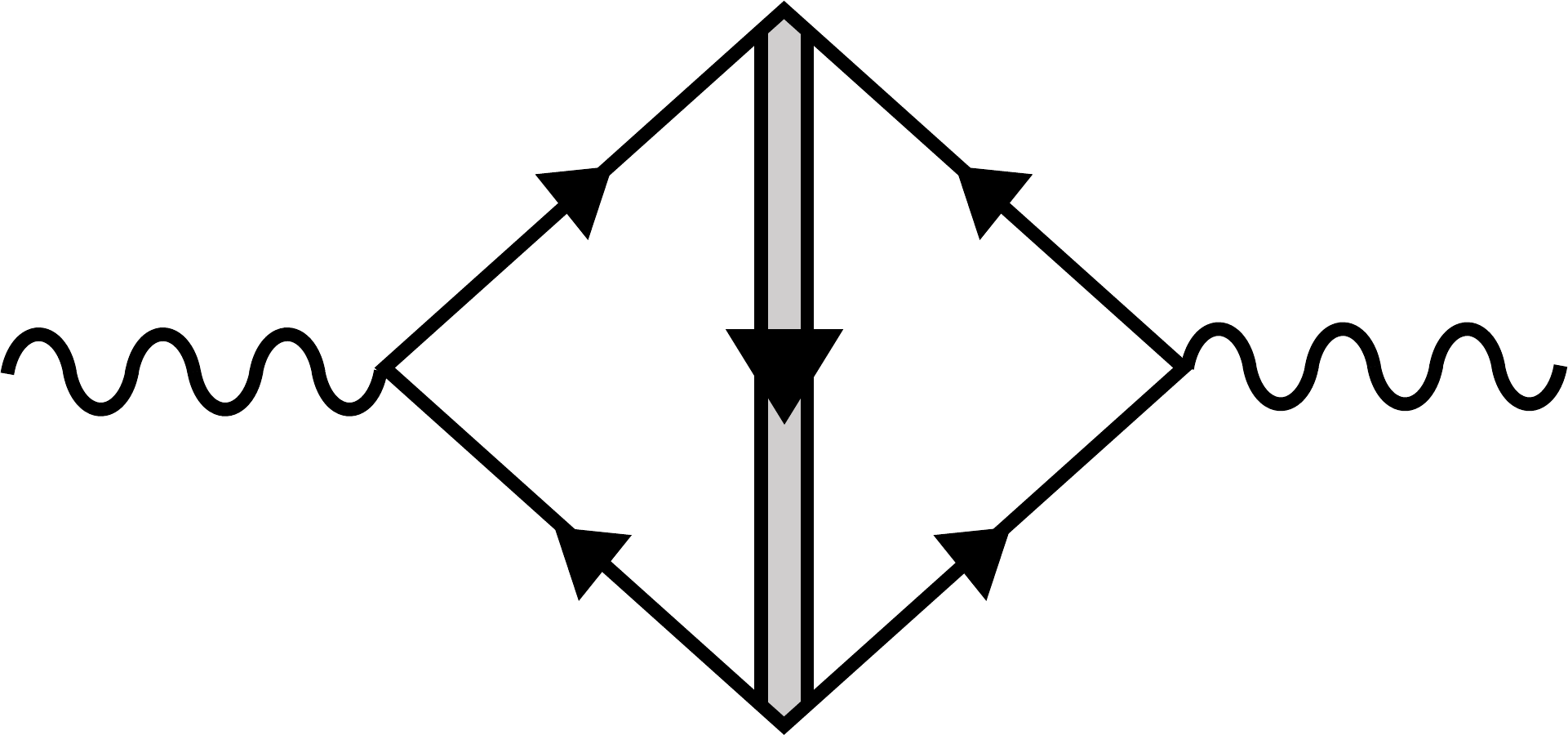}
       \subcaption{}
        \label{fig_MT}
      \end{minipage} \\ \\ 
      \begin{minipage}[t]{0.45\hsize}
        \centering
        \includegraphics[bb = 0 0 917 410, keepaspectratio, scale=0.083]{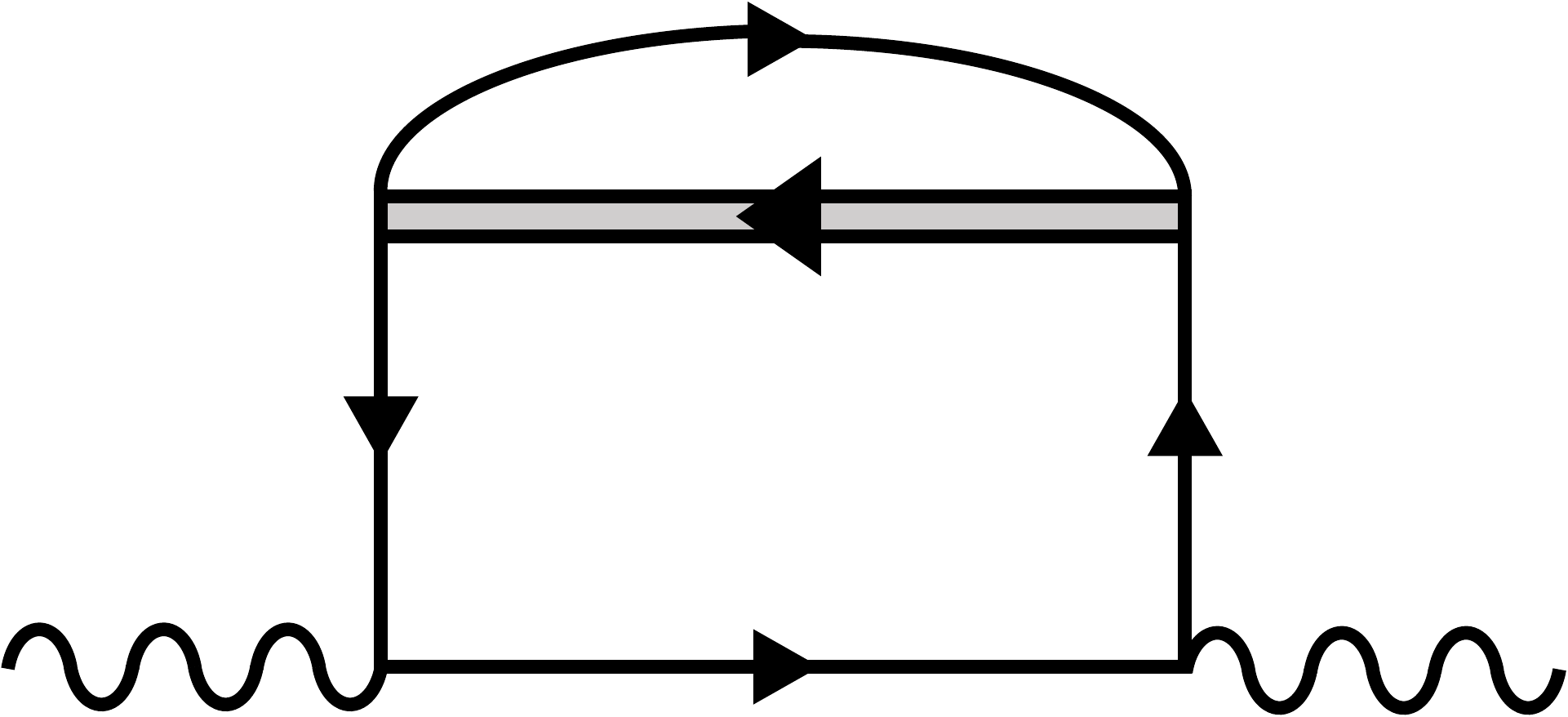}
        \subcaption{}
        \label{fig_DOS1}
      \end{minipage} &
      \begin{minipage}[t]{0.45\hsize}
        \centering
        \includegraphics[bb = 0 0 917 410, keepaspectratio, scale=0.083]{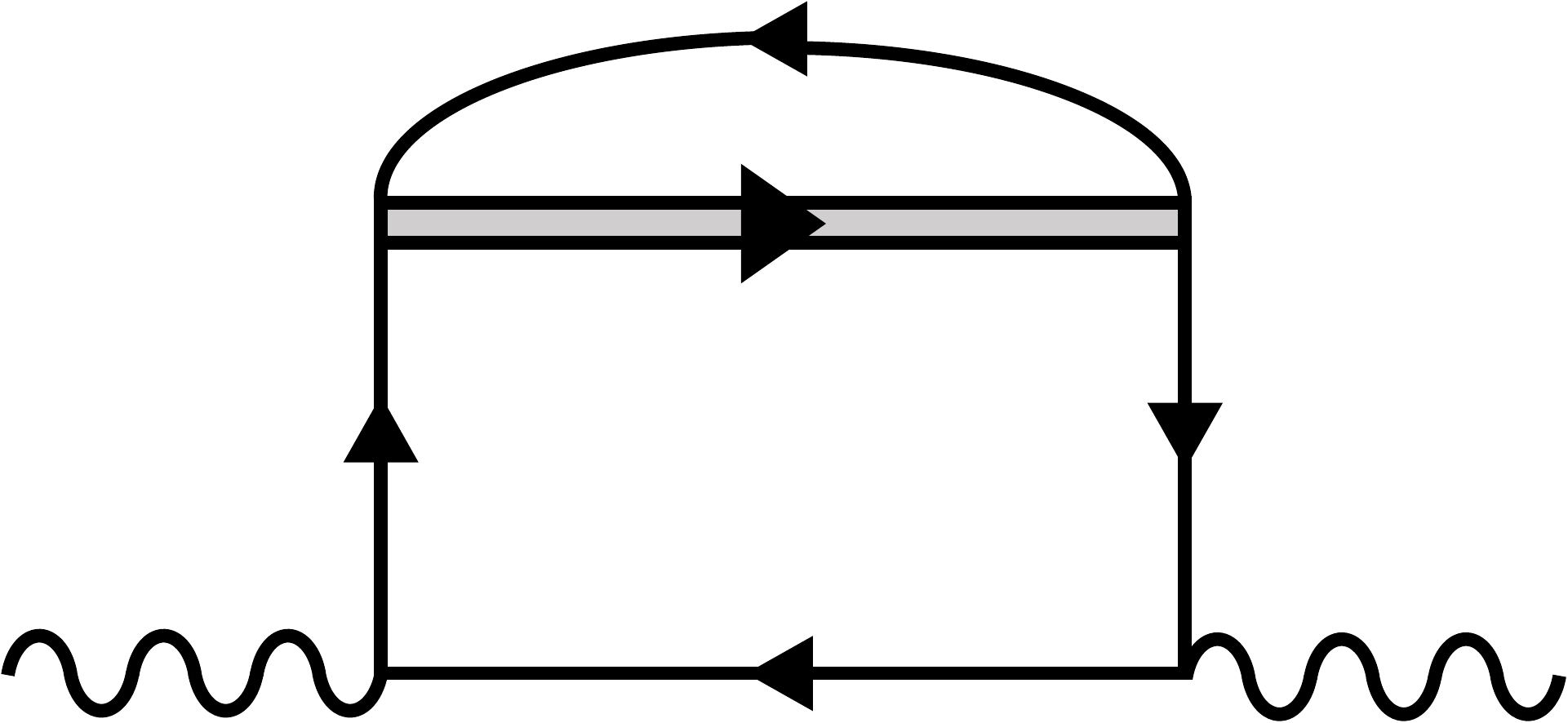}
        \subcaption{}
        \label{fig_DOS2}
      \end{minipage} 
    \end{tabular}
    \caption{Diagrammatic representations of the Aslamasov-Larkin (a),
      Maki-Thompson (b) and density of states (c,d) terms
      in Eqs.~(\ref{eq_AL})--(\ref{eq_DOS}).
      The double and wavy lines represent diquarks and
      photons, respectively.}
     \label{fig_selfenergy}
\end{figure}

We are now in a position to discuss the way how 
the effects of the diquark fluctuations are included in 
$\Pi^{R \mu\nu}(\bm{k}, \omega)$.
For that, we start from the one-loop diagram of the diquark propagator 
shown in Fig.~\ref{fig_potential}, which is the lowest-order
contribution of the diquark fluctuations to the thermodynamic potential.
The photon self-energy is then constructed by attaching 
electromagnetic vertices at two points of quark lines
in Fig.~\ref{fig_potential}.
This construction guarantees the WT identity $k_\nu \Pi^{R\mu\nu}(\bm{k},\omega)=0$.
This procedure leads to four types of diagrams 
shown in Fig.~\ref{fig_selfenergy},
which are called (a) Aslamasov-Larkin (AL)~\cite{AL:1968}, 
(b) Maki-Thompson (MT)~\cite{Maki:1968,Thompson:1968}
and (c,d) density of states (DOS) terms, respectively,
in the theory of metallic SC~\cite{book_Larkin}. 
The respective contributions to the photon self-energy
are denoted by $\tilde\Pi_{\rm AL}^{\mu\nu} (k)$, $\tilde\Pi_{\rm MT}^{\mu\nu} (k)$ and
$\tilde\Pi_{\rm DOS}^{\mu\nu} (k)$ in  the imaginary-time formalism, 
which are expressed as
\begin {align}
  \tilde\Pi_{\rm AL}^{\mu\nu} (k) &= 3
  \int_q
  \tilde\Gamma^\mu(q, q+k) \tilde\Xi(q+k) \tilde\Gamma^\nu(q+k, q) \tilde\Xi(q),
  \label{eq_AL}
  \\
  \tilde\Pi_{\rm MT}^{\mu\nu} (k) &= 3
  \int_q
  \tilde\Xi(q) \ \mathcal{R}_{\rm MT}^{\mu\nu}(q, k),
  \label{eq_MT}
  \\
  \tilde\Pi_{\rm DOS}^{\mu\nu} (k) &= 3
  \int_q
  \tilde\Xi(q) \ \mathcal{R}_{\rm DOS}^{\mu\nu}(q, k),
  \label{eq_DOS} 
\end {align}
respectively, 
where $q=(\bm{q}, i\nu_n)$ is the four momentum of a diquark field
and the overall coefficients $3$ come from three antisymmetric channels
of the diquark field.
The vertex functions $\tilde\Gamma^\mu(q, k)$, $\mathcal{R}_{\rm MT}^{\mu\nu} (q, k)$ and
$\mathcal{R}_{\rm DOS}^{\mu\nu} (q, k)$ in Eqs.~(\ref{eq_AL})--(\ref{eq_DOS}) are given by
\begin {align}
  &\tilde\Gamma^\mu (q, q+k) = 8(e_u+e_d)
  \int_p
  {\rm Tr} [\mathcal{G}_0 (p)\gamma^\mu\mathcal{G}_0 (p+k)\mathcal{G}_0 (q-p)],
  \label{eq_Gamma} \\
  &\mathcal{R}_{\rm MT}^{\mu\nu} (q, k) = 16e_u e_d \ \nonumber \\
  &\times
  \int_p
  {\rm Tr} [\mathcal{G}_0 (p)\gamma^\mu\mathcal{G}_0 (p+k)\mathcal{G}_0 (q-p-k) 
    \gamma^\nu\mathcal{G}_0 (q-p)],
  \label{eq_RMT} \\
  &\mathcal{R}_{\rm DOS}^{\mu\nu} (q, k) = 8(e^2_u+e^2_d) \ \nonumber \\
  &\times
  \int_p
  \Big\{ {\rm Tr} [\mathcal{G}_0 (p)\gamma^\mu\mathcal{G}_0 (p+k)\gamma^\nu \mathcal{G}_0 (p) \mathcal{G}_0 (q-p)] \nonumber \\
  &\qquad + {\rm Tr} [\mathcal{G}_0 (p) \gamma^\mu
    \mathcal{G}_0 (p-k) \gamma^\nu \mathcal{G}_0 (p) \mathcal{G}_0 (q-p)] \Big\}.
  \label{eq:RDOS}
\end {align}
where 
$e_u=2|e|/3$ and $e_d=-|e|/3$ are the electric charges of up and down quarks,
respectively, with the elementary charge $e$.

The total photon self-energy in imaginary time is given by
\begin {align}
\tilde\Pi^{\mu\nu} (k) &= \tilde\Pi_{\rm free}^{\mu\nu} (k) + \tilde\Pi_{\rm fluc}^{\mu\nu} (k), 
\label{eq_Pi_all} \\
\tilde\Pi_{\rm fluc}^{\mu\nu} (k) &= \tilde\Pi_{\rm AL}^{\mu\nu} (k)
+\tilde\Pi_{\rm MT}^{\mu\nu} (k)+\tilde\Pi_{\rm DOS}^{\mu\nu} (k),
\label{eq_Pi_fluc}
\end {align}
where $\tilde\Pi^{\mu\nu}_{\rm fluc}(k)$ denotes the modification of 
the self-energy due to the diquark fluctuations and
$\tilde\Pi^{\mu\nu}_{\rm free}(k)$ is that of the free quark system~\cite{book_Kapusta,book_LeBellac}.

\subsection{Vertices}
\label{sec:vertex}

The vertices (\ref{eq_Gamma})--(\ref{eq:RDOS}) satisfy the WT identities 
\begin {align}
  k_\mu \tilde\Gamma^\mu (q, q+k)
  =& e_\Delta \big({\cal Q}(q+k) - {\cal Q}(q) \big) 
  = e_\Delta \bigg( \frac{1}{\tilde\Xi (q+k)} - \frac{1}{\tilde\Xi (q)} \bigg) \ ,
  \label{eq_ALvertex_Ward} 
  \\
  k_\mu \mathcal{R}^{\mu\nu} (q, k) =& e_\Delta \big( \Gamma^\nu (q-k, q)-\Gamma^\nu (q, q+k) \big),
  \label{eq_MTvertex_Ward}
\end {align}
with $\mathcal{R}^{\mu\nu} (q, k) =  \mathcal{R}_{\rm MT}^{\mu\nu} (q, k) + \mathcal{R}_{\rm DOS}^{\mu\nu} (q, k)$
and $e_\Delta=e_u+e_d$ being the electric charge of diquarks.
Using Eqs.~(\ref{eq_ALvertex_Ward}), (\ref{eq_MTvertex_Ward}) and
$\tilde\Gamma^\nu (q, q+k)=\tilde\Gamma^\nu(q+k,q)$,
the WT identity of the photon self-energy
$k_\nu \tilde\Pi^{\mu\nu}_{\rm fluc}(k)=0$ is shown explicitly as 
\begin {align}
  k_\mu \tilde\Pi_{\rm fluc}^{\mu\nu} (k)
  =& ~k_\mu \tilde\Pi_{\rm AL}^{\mu\nu} (k) + k_\mu \big\{\tilde\Pi_{\rm MT}^{\mu\nu} (k)+\tilde\Pi_{\rm DOS}^{\mu\nu} (k) \big\} \nonumber \\
  =&-3 e_\Delta
  \int_q
  \big[ \tilde\Xi(q+k)-\tilde\Xi(q)\big] \tilde\Gamma^\nu(q, q+k)
  \nonumber \\
  &+3 e_\Delta
  \int_q
  \tilde\Xi(q) \big[ \tilde\Gamma^\nu(q-k, q) - \tilde\Gamma^\nu(q, q+k)\big] 
  \nonumber \\
  =&0.
  \label{eq:WT=0}
\end {align}

Since we adopt the TDGL approximation for $\Xi^R(\bm{k},\omega)$,
the vertices $\tilde\Gamma^\mu(q,q+k)$ and $\mathcal{R}^{\mu\nu}(q,k)$ have to be approximated
to satisfy Eqs.~(\ref{eq_ALvertex_Ward}) and (\ref{eq_MTvertex_Ward}) 
within this approximation.
From Eq.~(\ref{eq_softmodeapprox}) one finds
\begin{align}
  & [\tilde\Xi(q+k)]^{-1} - [\tilde\Xi(q)]^{-1} 
  \simeq c_0 i\nu_n + Q(\bm{q}+\bm{k},0) - Q(\bm{q},0)
  \nonumber \\
  &= c_0 i\nu_n + \frac{Q(\bm{q}+\bm{k},0)
    - Q(\bm{q},0)}{|\bm{q}+\bm{k}|^2 - |\bm{q}|^2}
            (|\bm{q}+\bm{k}|^2 - |\bm{q}|^2)
  \nonumber \\
  &= c_0 i\nu_n + Q_{(1)}(\bm{q}+\bm{k},\bm{q})~
  (2\bm{q}+\bm{k})\cdot \bm{k},
  \label{eq:Xi-Xi}
\end{align}
where $Q_{(1)}(\bm{q}_1,\bm{q}_2)
=(Q(\bm{q}_1,0) - Q(\bm{q}_2,0))/(|\bm{q}_1|^2 - |\bm{q}_2|^2)$
is finite in the limit $|\bm{q}_1-\bm{q}_2|\to0$
because $Q(\bm{q},\omega)$ is a function of $|\bm{q}|^2$.
Substituting Eq.~(\ref{eq:Xi-Xi}) into Eq.~(\ref{eq_ALvertex_Ward})
and requiring the analyticity of $\tilde\Gamma^\mu(q,q+k)$
at $\omega=|\bm{k}|=0$ one finds that $\tilde\Gamma^0(q,q+k) = e_\Delta c_0$ and 
\begin {align}
  \tilde\Gamma^i (q, q+k) 
  = - e_\Delta  Q_{(1)}(\bm{q}+\bm{k},\bm{q}) (2q+k)^i,
  \label{eq_ALvertex_approx}
\end{align}
are choices that satisfy Eq.~(\ref{eq_ALvertex_Ward}),
where $i = 1, 2, 3$.
One can also obtain forms of $\mathcal{R}^{\mu\nu}(q,k)$ satisfying
Eq.~(\ref{eq_MTvertex_Ward}) with Eq.~(\ref{eq_ALvertex_approx})
in a similar manner, which, however, are not shown explicitly
since 
they turn out unnecessary in this study as discussed below.
These vertices with Eq.~(\ref{eq_softmodeapprox}) satisfy
the WT identity of $\tilde\Pi^{\mu\nu}(k)$.
It should be warned, however, that the uniqueness of the choice of 
Eq.~(\ref{eq_ALvertex_approx}) holds 
only in the lowest order of $\omega$ and $|\bm{k}|^2$,
and hence the non-uniqueness may affect
the final result in the high energy region.

\subsection{Dilepton production rate}

In the above construction of $\tilde\Gamma^\mu (q, q+k)$ and
$\mathcal{R}^{\mu\nu}(q,k)$, 
the spatial components of these vertices are real.
This fact greatly simplifies the analytic continuation from
$\tilde\Pi^{ij}_{\rm fluc}(k)$ to $\Pi^{R ij}_{\rm fluc}(\bm{k},\omega)$.
From the reality of $\mathcal{R}^{ij}(q,k)$ it is also shown that 
${\rm Im}[\Pi^{Rij}_{\rm MT}(\bm{k},\omega)+\Pi^{Rij}_{\rm DOS}(\bm{k},\omega)]=0$~\cite{book_Larkin},
which means that the spatial components 
${\rm Im}\Pi^{R ij}_{\rm fluc}(\bm{k},\omega)$ only come from
$\Pi_{\rm AL}^{R ij}(\bm{k},\omega)$,
while the temporal component ${\rm Im}\Pi^{R00}_{\rm fluc}(\bm{k},\omega)$ is
given by the sum of AL, MT and DOS terms.
The temporal component, however, is obtained from the
spatial ones using the WT identity
\begin {align}
  \tilde\Pi^{00} (k) = \frac{\bm{k}^2}{(i\nu_l)^2} \tilde\Pi^{11}(k),
  \label{eq_WT}
\end {align}
with $k=(i\nu_l, |\bm{k}|, 0, 0)$.
One then finds that $g_{\mu\nu} \tilde\Pi_{\rm fluc}^{\mu\nu}(k)$
in Eq.~(\ref{eq_RATE_kom}) is obtained only
from $\Pi_{\rm AL}^{R ij}(\bm{k},\omega)$ as 
\begin {align}
&g_{\mu\nu} \tilde\Pi_{\rm fluc}^{\mu\nu}(k)
= \frac{\bm{k}^2}{(i\nu_l)^2} \tilde\Pi_{\rm AL}^{11}(k) - \sum_{i=1}^{3} \tilde\Pi_{\rm AL}^{ii}(k) 
\nonumber \\
&=3 \int \frac{d^3\bm{q}}{(2\pi)^3} 
\bigg[ \frac{\bm{k}^2}{(i\nu_l)^2} \big(\tilde\Gamma^1 (q, q+k) \big)^2 
-\sum_i \big(\tilde\Gamma^i (q, q+k) \big)^2 \bigg] 
\nonumber \\
&\qquad\qquad\times
\oint_C \frac{dq_0}{2\pi i} \frac{\coth \frac{q_0}{2T}}{2}
\tilde\Xi(\bm{q}+\bm{k}, q_0+i\nu_l)\tilde\Xi(\bm{q}, q_0)
\label{eq:gPi'} \\
&=3 \int \frac{d^3\bm{q}}{(2\pi)^3} 
\bigg[ 
\frac{\bm{k}^2}{(i\nu_l)^2} \big(\tilde\Gamma^1 (q, q+k) \big)^2 - \sum_i \big(\tilde\Gamma^i (q, q+k) \big)^2 
\bigg] 
\nonumber \\
&\qquad\quad \times \bigg\{
P \int \frac{d \omega '}{2\pi i} \frac{{\rm coth} \frac{\omega '}{2T}}{2}
\Xi^R(\bm{q}+\bm{k}, \omega '+i\nu_l) \Xi^R(\bm{q}, \omega ') 
\nonumber \\
&\qquad\quad\ \ - P \int \frac{d \omega '}{2\pi i} \frac{{\rm coth} \frac{\omega '}{2T}}{2}
\Xi^R(\bm{q}+\bm{k}, \omega '+i\nu_l) \Xi^A(\bm{q}, \omega ') 
\nonumber \\
&\qquad\quad\ \  + P \int \frac{d \omega '}{2\pi i} \frac{{\rm coth} \frac{\omega '}{2T}}{2}
\Xi^R(\bm{q}+\bm{k}, \omega ') \Xi^A(\bm{q}, \omega '-i\nu_l) 
\nonumber \\
&\qquad\quad\ \  - P \int \frac{d \omega '}{2\pi i} \frac{{\rm coth} \frac{\omega '}{2T}}{2}
\Xi^A(\bm{q}+\bm{k}, \omega ') \Xi^A(\bm{q}, \omega '-i\nu_l) 
\bigg\},
\label{eq:gPi}
\end {align}
where the contour $C$ in Eq.~(\ref{eq:gPi'}) surrounds
the poles of $\coth (q_0/2T)$
and $\Xi^A(\bm{k}, \omega) = \tilde\Xi(k)|_{i\nu_l\to\omega-i\eta}$ is
the advanced T-matrix.
The far right-hand side Eq.~(\ref{eq:gPi})
is obtained after deforming the contour $C$
avoiding the cut in $\tilde\Xi(q)$ on the real axis~\cite{book_Larkin}.
By taking the analytic continuation
$i\nu_l \rightarrow \omega + i\eta$ and using
$\Xi^A(\bm{k}, \omega) =[\Xi^R(\bm{k}, \omega)]^*$, 
we obtain
\begin {align}
&g_{\mu\nu} {\rm Im} \Pi_{\rm fluc}^{R\mu\nu}(\bm{k}, \omega) 
  =3 e_\Delta^2 \int_{-2\Lambda-2\mu}^{2\Lambda-2\mu} \frac{d \omega '}{2\pi}
  \int \frac{d^3\bm{q}}{(2\pi)^3} {\rm coth} \frac{\omega '}{2T} 
\nonumber \\
&\times 
\big(Q_{(1)}(\bm{q}+\bm{k},\bm{q})\big)^2
\Bigg[ \Bigg(\frac{(\bm{q}+\bm{k})^2-\bm{q}^2}{\omega}\Bigg)^2 - (2\bm{q}+\bm{k})^2 \Bigg]
\nonumber \\
&\times 
{\rm Im} \Xi^R (\bm{q}+\bm{k}, \omega ') 
\Big\{
{\rm Im} \Xi^R(\bm{q}, \omega '+\omega) - {\rm Im} \Xi^R(\bm{q}, \omega '-\omega)
\Big\}.
\label{eq:ImPi}
\end {align}
To deal with the momentum integral in Eq.~(\ref{eq:ImPi}),
we introduce the ultraviolet cutoff with the same procedure
as in Ref.~\cite{Kitazawa:2005vr}.
The DPR is obtained by substituting this result
into Eq.~(\ref{eq_RATE_kom}).

\begin{figure*}[tbp]
\begin{tabular}{ccc}
      \begin{minipage}[t]{0.31\hsize}
        \centering
        \includegraphics[bb = 0 0 917 424, keepaspectratio, scale=0.31]{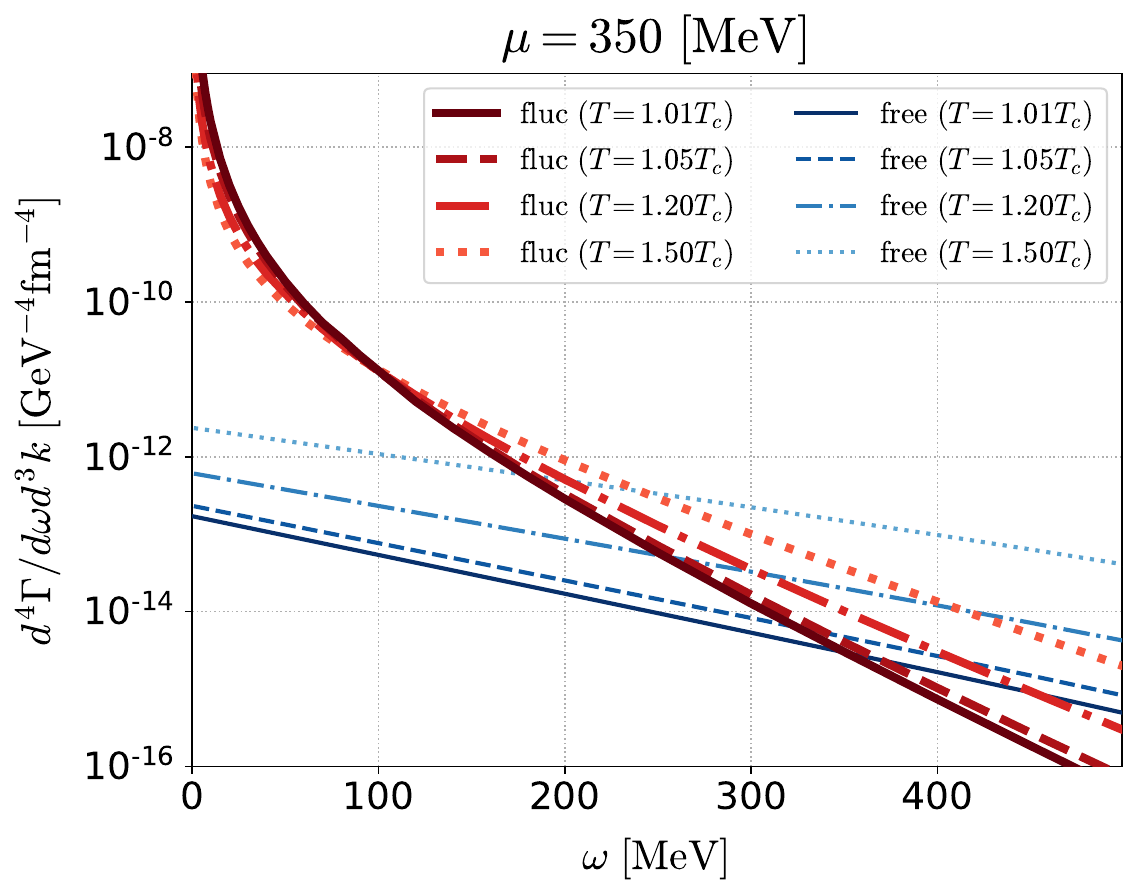}
      \end{minipage} &
      \begin{minipage}[t]{0.31\hsize}
        \centering
        \includegraphics[bb = 0 0 918 428, keepaspectratio, scale=0.31]{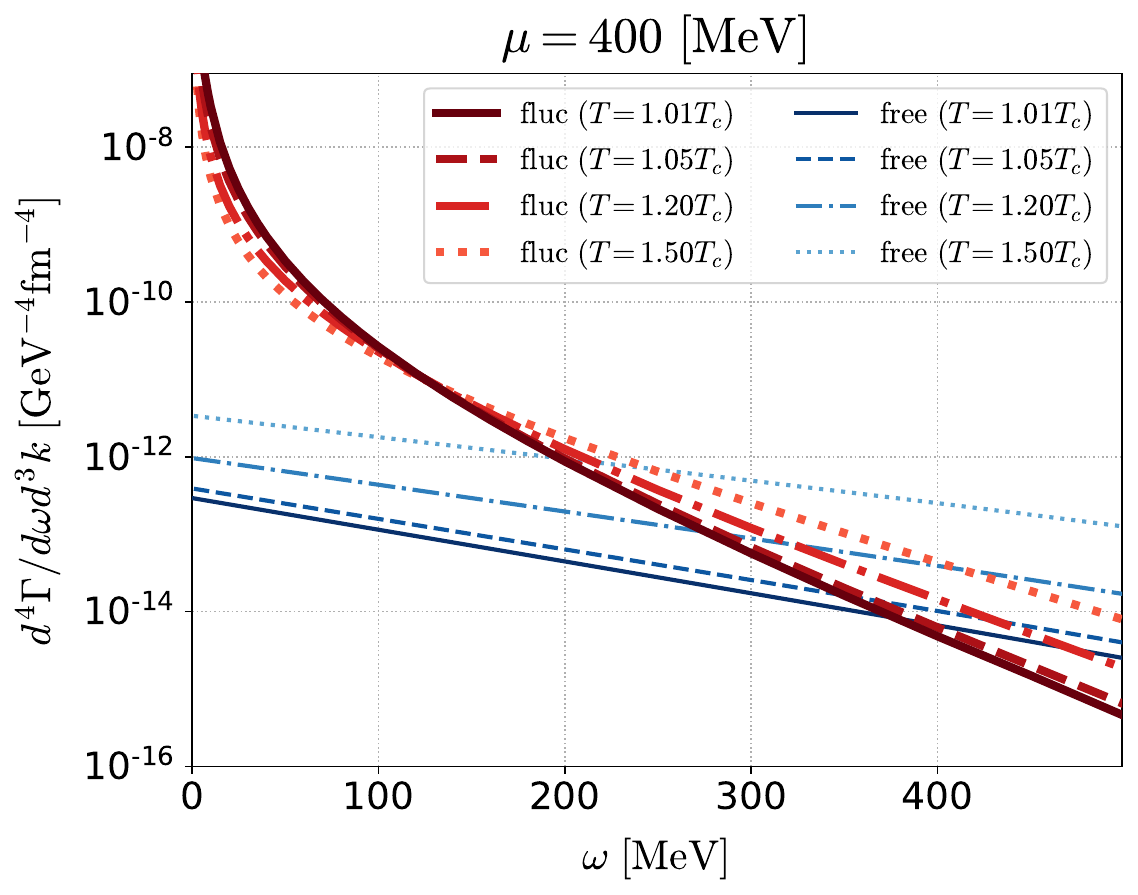}
      \end{minipage} &
      \begin{minipage}[t]{0.31\hsize}
        \centering
        \includegraphics[bb = 0 0 917 410, keepaspectratio, scale=0.31]{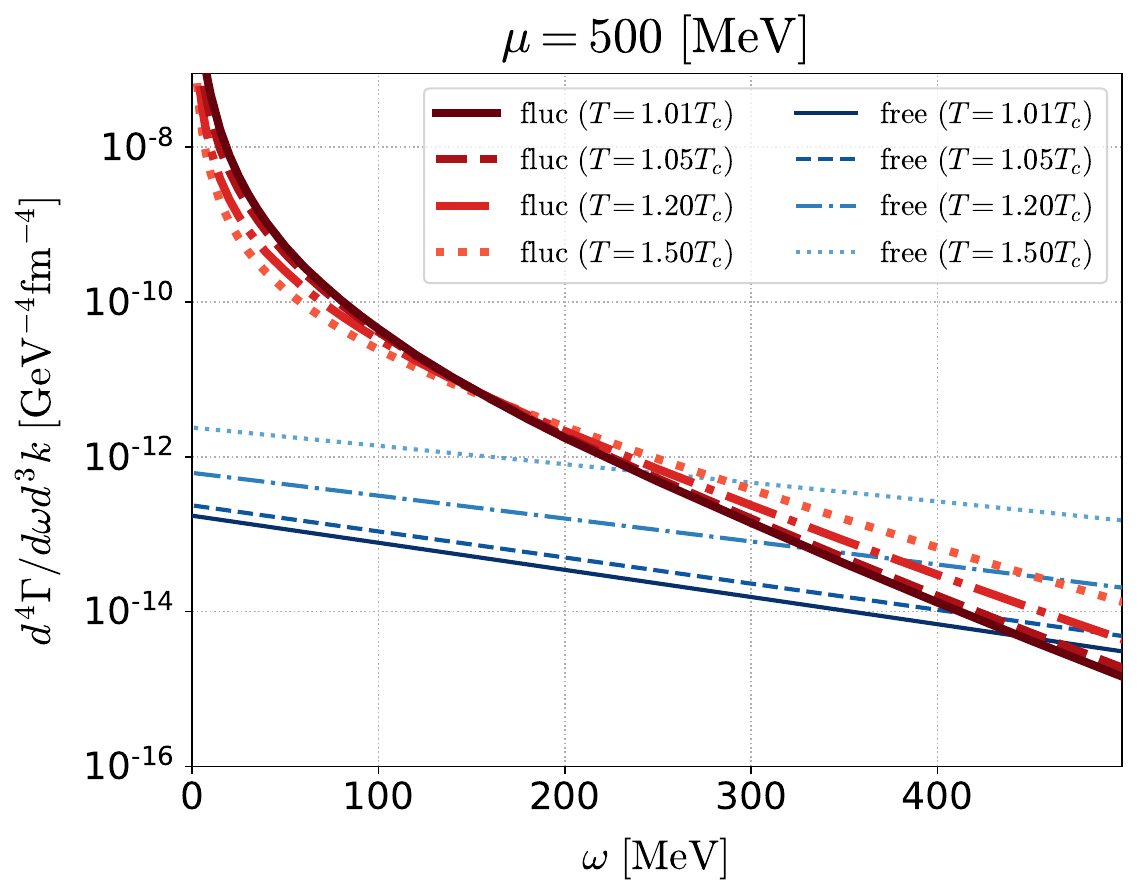}
      \end{minipage} 
\end{tabular}
\caption{
Dilepton production rates per unit energy and momentum 
$d^4 \Gamma / d\omega d^3k$
at $\bm{k}=0$ for several values of $T/T_c$ with
$\mu=350$~MeV (left), $400$~MeV (middle)
and $500$~MeV (right) and $G_{\rm C}=0.7G_{\rm S}$. 
The thick-red (thin-blue) lines show the contribution of 
$\tilde\Pi_{\rm fluc}^{\mu\nu}(k)$ ($\tilde\Pi_{\rm free}^{\mu\nu}(k))$.
}
\label{fig_RATE_0om}
\end{figure*}

\section{Numerical results}

In Fig.~\ref{fig_RATE_0om}, we show the numerical results of
the production rate $d^4\Gamma / d^4k$ 
per unit energy and momentum at $\bm{k}=\bm{0}$
calculated with use of the photon self-energy Eq.~(\ref{eq_Pi_all})
and Eq.~(\ref{eq:ImPi})
for various values of $T$ and $\mu$ at $G_{\rm C} = 0.7G_{\rm S}$.
The thick lines show the contribution of diquark fluctuations
obtained from $\Pi_{\rm fluc}^{R\mu\nu}(\bm{k},\omega)$,
while the thin lines are the results for the free quark gas.
The total rate is given by the sum of these two contributions.
The figure shows that the production rate 
is enhanced so much by the diquark fluctuations 
that it greatly exceeds that of the free quarks
in the low energy region $\omega\lesssim300$~MeV.
The enhancement is more pronounced as $T$ is lowered toward $T_c$,
while the enhancement at $\omega\simeq200$~MeV
is observed up to $T \simeq 1.5T_c$.
The figure also shows that the contribution of diquark fluctuations is
more enhanced as $\mu$ becomes larger.
This behavior is understood as the effect of the
larger Fermi surface for larger $\mu$.

\begin{figure}[tbp]
\centering
\includegraphics[width=0.43\textwidth]{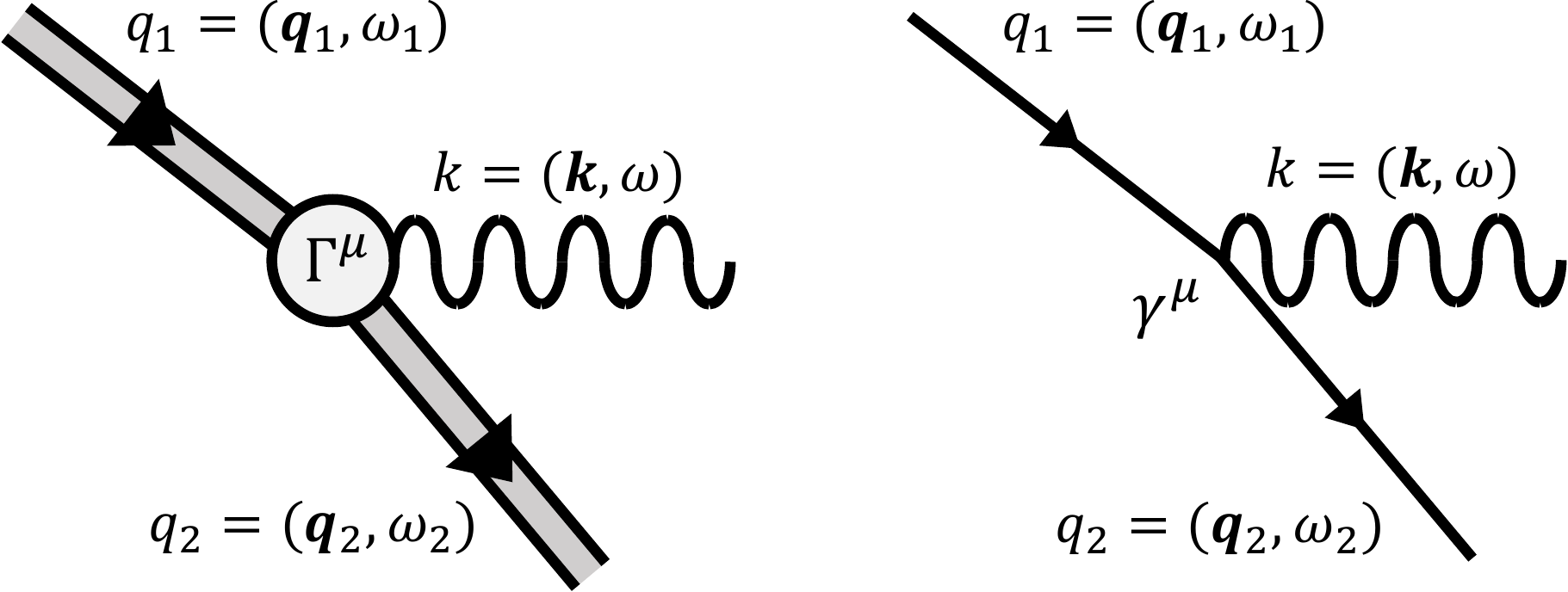}
\caption{Diagrams representing the processes of a virtual photon production.
}
\label{fig_mechanism}
\end{figure}

It is found worth scrutinizing the underlying mechanism 
of the low-energy enhancement of the production mechanism
of virtual photons. 
Although it is rather natural that $d^4\Gamma / d^4k$ is enhanced
in the  low energy region since the virtual photons are emitted from the
soft collective modes,
their pronounced effects on the production of virtual photons
in the {\em time-like} region deserves an elucidation
since the soft mode has a dominant strength 
in the {\em space-like} region as shown in Fig.~\ref{fig_DSF}.
In our formalism, the virtual photons are dominantly emitted
through the process obtained by cutting Fig.~\ref{fig_selfenergy} (a),
i.e. the scattering of diquarks
shown in the left panel of Fig.~\ref{fig_mechanism}.
In this process, energy-momentum of the virtual photon
$k=(\bm{k},\omega)$ can be time-like, $\omega>\vert \bm{k}\vert$,
since 
the absolute value of the momentum $\bm{k}=\bm{q}_1-\bm{q}_2$ can be taken
arbitrarily small keeping $\omega=\omega_1-\omega_2$ finite.
This kinematics is contrasted to the scattering of massless quarks
shown in the right panel of Fig.~\ref{fig_mechanism}, in which 
the produced virtual photon is always in the space-like region
$\omega<\vert \bm{k}\vert$.
However, $\omega=\omega_1-\omega_2$ of a virtual photon is restricted
to small values due to the small energies $\omega_1$ and $\omega_2$
of diquarks.
The sharp peak of $d^4\Gamma / d^4k$ in Fig.~\ref{fig_RATE_0om}
is understood in this way.

\begin{figure*}[tbp]
\begin{tabular}{ccc}
\begin{minipage}[t]{0.32\hsize}
\centering
\includegraphics[keepaspectratio, scale=0.47]{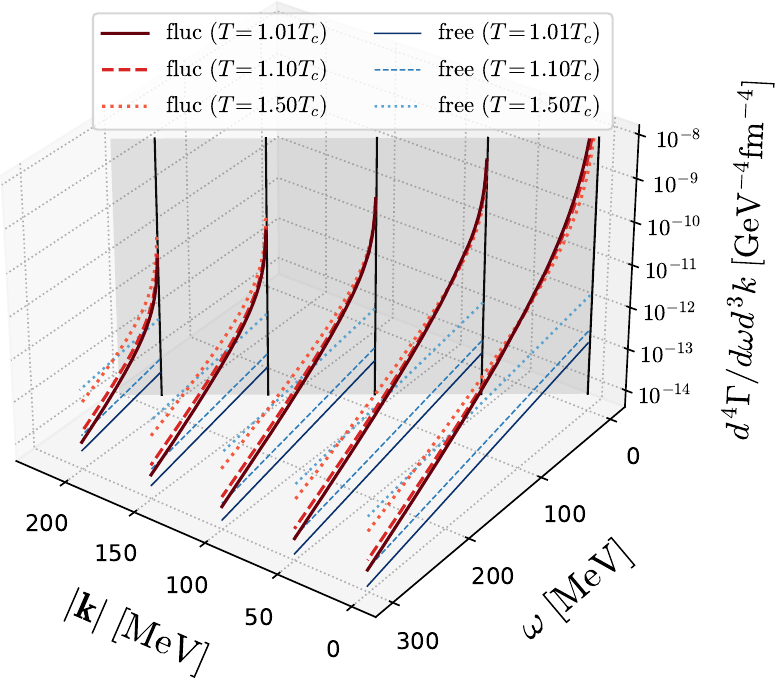}
\end{minipage} &
\begin{minipage}[t]{0.32\hsize}
\centering
\includegraphics[keepaspectratio, scale=0.3]{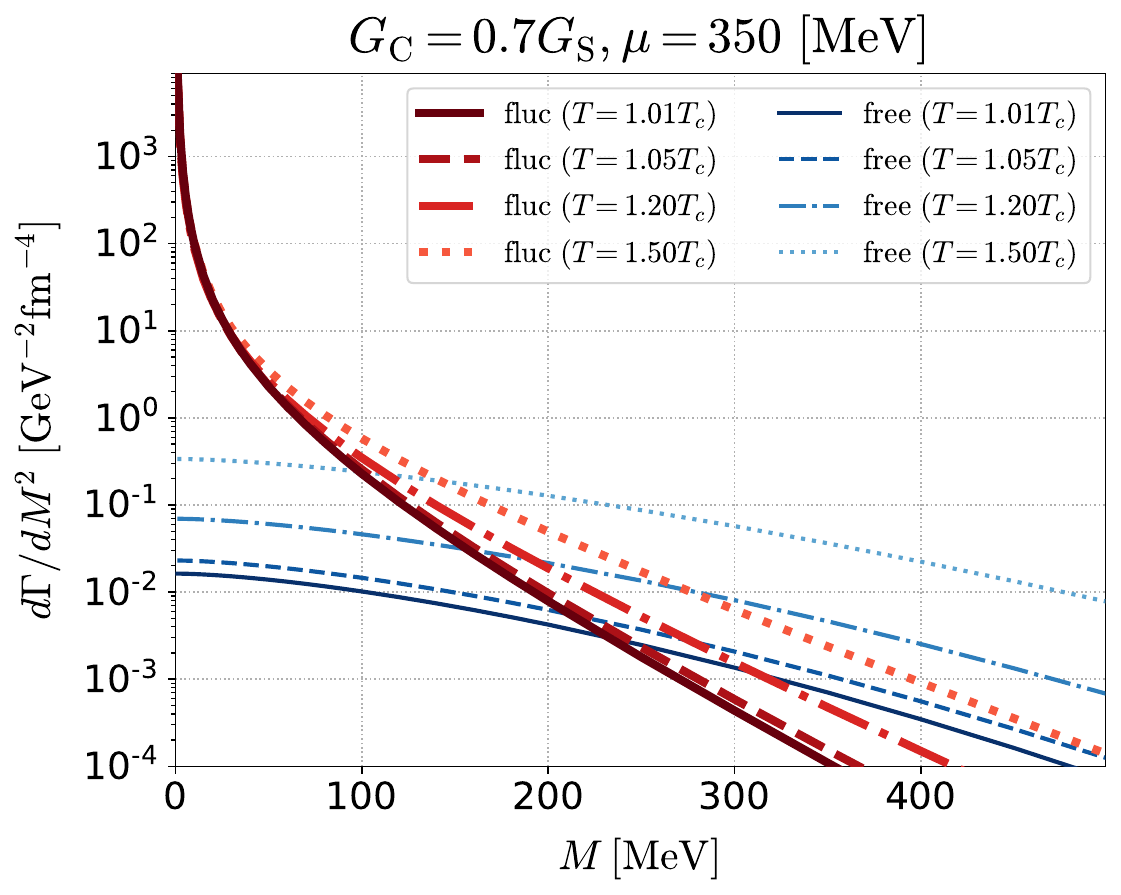}
\end{minipage} 
\begin{minipage}[t]{0.32\hsize}
\centering
\includegraphics[keepaspectratio, scale=0.3]{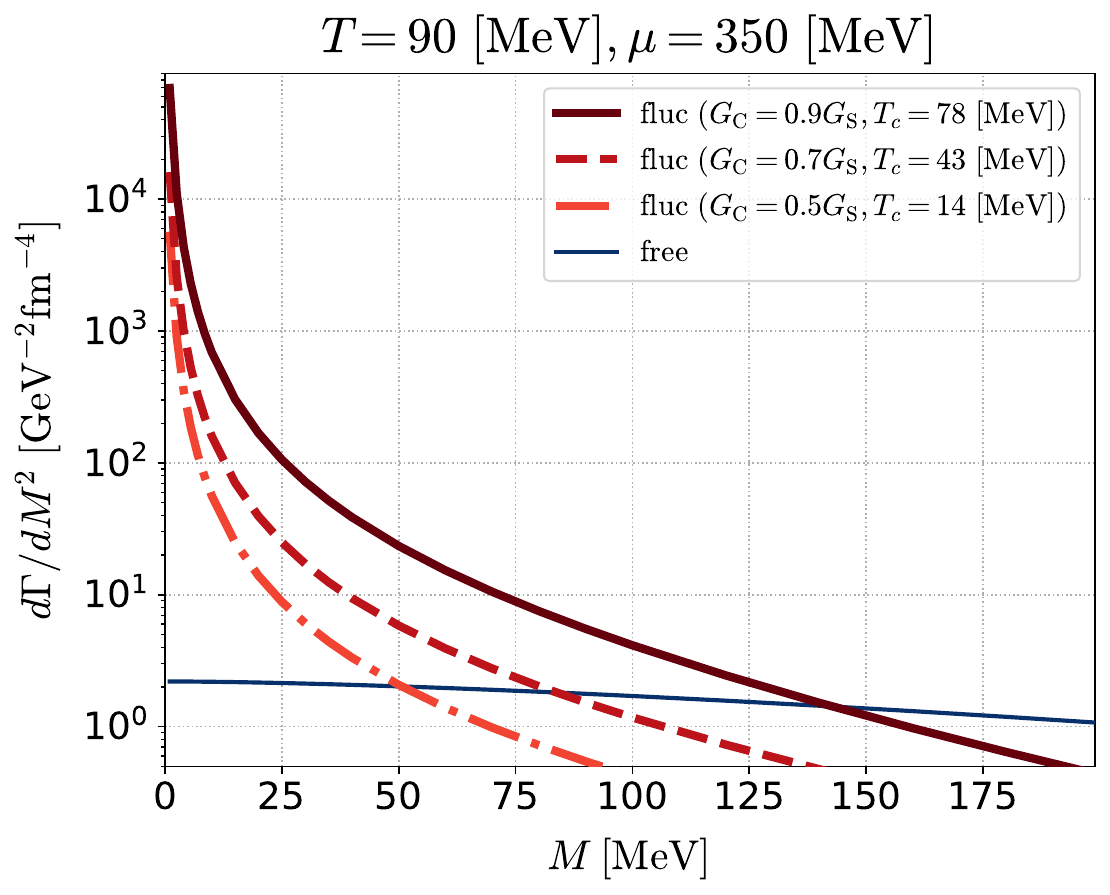}
\end{minipage} 
\end{tabular}
\caption{
  \textbf{Left}: Dilepton production rates $d^4 \Gamma / d\omega d^3k$
  as a function of $\omega$ and $|\bm{k}|$
  for $\mu=350$~MeV and $T=1.01~T_c$, $1.1~T_c$ and $1.5~T_c$ 
  at $G_{\rm C}=0.7G_{\rm S}$.
  The gray surface shows the light-cone. 
  \textbf{Middle}: 
  The invariant-mass spectrum $d\Gamma/dM^2$ for several values of $T/T_c$
  at $G_{\rm C}=0.7G_{\rm S}$ and $\mu =350$~MeV.
  \textbf{Right}: The invariant-mass spectrum $d\Gamma/dM^2$
  at $(T,\mu) =(90,350)$~MeV 
  for $G_{\rm C}=0.9G_{\rm S}$ (solid), $0.7G_{\rm S}$ (dashed) and $0.5G_{\rm S}$ (dash-dotted). 
}
\label{fig_RATE}
\end{figure*}

To have more detailed properties of the enhancement of DPR,
we show in the far left panel of Fig.~\ref{fig_RATE}
a three-dimensional plot of DPR 
in the $\omega$--$|\bm{k}|$ plane for several values of $T$
at $\mu=350$~MeV and $G_{\rm C}=0.7G_{\rm S}$.
We see that the DPR is enhanced strongly 
around the origin in the $\omega$--$|\bm{k}|$ plane, 
and the larger $\omega$ and/or  $|\bm{k}|$, the smaller the DPR. 
This behavior is in accordance with the mechanism explained above.

In the HIC experiments, the dilepton production rate is usually measured
as a function of the invariant mass, $M$,
\begin {align}
  \frac{d\Gamma}{dM^2}
  = \int d^3k \frac{1}{2\omega} \frac{d^4\Gamma}{d^4k} \bigg|_{\omega=\sqrt{k^2+M^2}}\ .
  \label{eq:dGdM}
\end {align}
In the middle panel of Fig.~\ref{fig_RATE}, we show Eq.~(\ref{eq:dGdM})
for several values of $T/T_c$ at $G_{\rm C}=0.7G_{\rm S}$ and $\mu = 350$~MeV.
One sees that the enhancement due to diquark fluctuations
is observed in the low invariant-mass region 
$M < (150 -200)$~MeV up to $T\simeq1.5T_c$.
The little $T$ dependence of DPR seen in the far low region of $M$
may be understood as a result of an accidental 
cancellation between the enhanced spectral function due to 
the soft mode and the kinematical thermal effect:
The sharp enhancement of the former 
at low energy-momentum near $T_c$ decreases while
the creation probability due to the thermal effect 
increases as $T$ goes high.
The contribution of the diquark fluctuations is relatively suppressed
for higher $T$ as the contribution of free quarks becomes larger.

Finally, shown in the right panel of Fig.~\ref{fig_RATE} is
$d\Gamma/dM^2$ at fixed $(T,\mu)=(90,350)$~MeV
(the cross symbol in Fig.~\ref{fig_phase})
for several values of $G_{\rm C}$.
The panel shows that the production rate is more enhanced
for larger $G_{\rm C}$ and $T_c$.
For $G_{\rm C}=0.9G_{\rm S}$ ($T_c\simeq78$~MeV),
the production rate from the diquark fluctuations exceeds those of
the free quarks for $M\lesssim100$~MeV.

\section{Discussions}

In this Letter, we have investigated the effect of diquark
fluctuations on the DPR 
near but above the critical temperature of the 2SC.
The contribution of the diquark fluctuations were 
taken into account through the AL, MT and DOS terms
in the photon self-energy.
We have found that the dilepton production rate 
is strongly enhanced in comparison with the free-quark gases 
in the low energy and low invariant-mass regions near $T_c$
up to $T\simeq1.5T_c$ reflecting the formation of the diquark soft mode
associated with the phase transition to 2SC.

We would say that it should be rewarding to try to make 
an experimental measurement of dileptons in that far low-mass region  
and examine the possible enhancement of the DPR in the HIC; 
if the enhancement is confirmed, it may possibly give an experimental 
evidence of strong diquark correlations, which lead to
the phase transition to CSC in dense quark matter.
Moreover, it is to be noted that the
DPR with vanishing energy/momentum is
directly related to 
the electric conductivity, as is evident 
from the fact that the AL, MT and DOS terms in condensed 
matter physics are responsible for the anomalous
enhancement of the electric conductivity (paraconductivity)
in metals above $T_c$ but in the close 
vicinity of the superconducting phase.

There are, however, many issues to be resolved
for making the measurements meaningful, in the 
sense that it can help in revealing the significance of the
diquark fluctuations prior to the phase transition to the 2SC in the 
dense matter.
Since observed yield of the dilepton production in the HIC is
a superposition of those with various origins in the space-time history,
we need to `disentangle' the observed total yield into 
those with the respective origins.
For that, it is necessary to quantitatively estimate 
the residence time around the phase boundary of the CSC, say,
with resort to dynamical models~\cite{Nara:2021fuu}.
Even when some enhancement of the DPR in the 
low $M$ region is identified, it is to be noted that
it may have come from a different  mechanism
due to medium effects~\cite{Rapp:2009yu,Laine:2013vma,Ghiglieri:2014kma}.
A comparison of our results with these effects 
constitutes future projects.

The experimental measurement of the DPR in the 
relevant low invariant-mass region $M\lesssim200$~MeV is not an easy task
because di-electrons, which are, among dileptons, only available 
in this energy range, are severely contaminated by the Dalitz decays, and
high-precision measurements both of $d\Gamma/dM^2$ and hadron spectrum
are necessary to extract interesting medium effects.
Despite these challenging requirements, 
it is encouraging
that the future HIC programs in GSI-FAIR, NICA-MPD and J-PARC-HI
are designed to carry out high-precision experiments~\cite{Galatyuk:2019lcf},
and also that new technical developments are vigorously being 
made~\cite{Adamova:2019vkf}.

Finally, we  remark that such an effort to reveal the significance of the 
enhanced diquark correlations in the hot and dense matter should also 
give some clue to
the modern development 
of hadron physics where 
possible diquark correlations in  hadron structures are
one of the hot topics~\cite{Barabanov:2020jvn}.

\section*{Acknowledgements}

The authors thank Naoki Yamamoto for his critical comments. 
T.~N. thanks JST SPRING (Grant No.~JPMJSP2138) and
Multidisciplinary PhD Program for Pioneering Quantum Beam Application.
This work was supported by JSPS KAKENHI 
(Grants No.~JP19K03872, No.~JP90250977, and No.~JP10323263).

\bibliographystyle{elsarticle-num} 
\bibliography{reference.bib}

\begin{thebibliography}{10}
\expandafter\ifx\csname url\endcsname\relax
  \def\url#1{\texttt{#1}}\fi
\expandafter\ifx\csname urlprefix\endcsname\relax\def\urlprefix{URL }\fi
\expandafter\ifx\csname href\endcsname\relax
  \def\href#1#2{#2} \def\path#1{#1}\fi

\bibitem{Fukushima:2010bq}
K.~Fukushima, T.~Hatsuda, {The phase diagram of dense QCD}, Rept. Prog. Phys.
  74 (2011) 014001.
\newblock \href {http://arxiv.org/abs/1005.4814} {\path{arXiv:1005.4814}},
  \href {https://doi.org/10.1088/0034-4885/74/1/014001}
  {\path{doi:10.1088/0034-4885/74/1/014001}}.

\bibitem{Fukushima:2013rx}
K.~Fukushima, C.~Sasaki, {The phase diagram of nuclear and quark matter at high
  baryon density}, Prog. Part. Nucl. Phys. 72 (2013) 99--154.
\newblock \href {http://arxiv.org/abs/1301.6377} {\path{arXiv:1301.6377}},
  \href {https://doi.org/10.1016/j.ppnp.2013.05.003}
  {\path{doi:10.1016/j.ppnp.2013.05.003}}.

\bibitem{Asakawa:1989bq}
M.~Asakawa, K.~Yazaki, {Chiral Restoration at Finite Density and Temperature},
  Nucl. Phys. A 504 (1989) 668--684.
\newblock \href {https://doi.org/10.1016/0375-9474(89)90002-X}
  {\path{doi:10.1016/0375-9474(89)90002-X}}.

\bibitem{Barducci:1989wi}
A.~Barducci, R.~Casalbuoni, S.~De~Curtis, R.~Gatto, G.~Pettini, {Chiral
  Symmetry Breaking in {QCD} at Finite Temperature and Density}, Phys. Lett. B
  231 (1989) 463--470.
\newblock \href {https://doi.org/10.1016/0370-2693(89)90695-3}
  {\path{doi:10.1016/0370-2693(89)90695-3}}.

\bibitem{Kitazawa:2002jop}
M.~Kitazawa, T.~Koide, T.~Kunihiro, Y.~Nemoto, {Chiral and color
  superconducting phase transitions with vector interaction in a simple model},
  Prog. Theor. Phys. 108~(5) (2002) 929--951, [Erratum: Prog.Theor.Phys. 110,
  185--186 (2003)].
\newblock \href {http://arxiv.org/abs/hep-ph/0207255}
  {\path{arXiv:hep-ph/0207255}}, \href {https://doi.org/10.1143/PTP.108.929}
  {\path{doi:10.1143/PTP.108.929}}.

\bibitem{Stephanov:1999zu}
M.~A. Stephanov, K.~Rajagopal, E.~V. Shuryak, {Event-by-event fluctuations in
  heavy ion collisions and the QCD critical point}, Phys. Rev. D 60 (1999)
  114028.
\newblock \href {http://arxiv.org/abs/hep-ph/9903292}
  {\path{arXiv:hep-ph/9903292}}, \href
  {https://doi.org/10.1103/PhysRevD.60.114028}
  {\path{doi:10.1103/PhysRevD.60.114028}}.

\bibitem{Galatyuk:2019lcf}
T.~Galatyuk, {Future facilities for high $\mu_B$ physics}, Nucl. Phys. A 982
  (2019) 163--169.
\newblock \href {https://doi.org/10.1016/j.nuclphysa.2018.11.025}
  {\path{doi:10.1016/j.nuclphysa.2018.11.025}}.

\bibitem{Kojo:2020ztt}
T.~Kojo, D.~Hou, J.~Okafor, H.~Togashi, {Phenomenological QCD equations of
  state for neutron star dynamics: Nuclear-2SC continuity and evolving
  effective couplings}, Phys. Rev. D 104~(6) (2021) 063036.
\newblock \href {http://arxiv.org/abs/2012.01650} {\path{arXiv:2012.01650}},
  \href {https://doi.org/10.1103/PhysRevD.104.063036}
  {\path{doi:10.1103/PhysRevD.104.063036}}.

\bibitem{Cierniak:2021knt}
M.~Cierniak, D.~Blaschke, {Hybrid neutron stars in the mass-radius diagram},
  Astron. Nachr. 342~(5) (2021) 819--825.
\newblock \href {http://arxiv.org/abs/2106.06986} {\path{arXiv:2106.06986}},
  \href {https://doi.org/10.1002/asna.202114000}
  {\path{doi:10.1002/asna.202114000}}.

\bibitem{Kojo:2021wax}
T.~Kojo, G.~Baym, T.~Hatsuda, {QHC21 equation of state of neutron star matter -
  in light of 2021 NICER data} (11 2021).
\newblock \href {http://arxiv.org/abs/2111.11919} {\path{arXiv:2111.11919}}.

\bibitem{Alford:2007xm}
M.~G. Alford, A.~Schmitt, K.~Rajagopal, T.~Sch\"afer, {Color superconductivity
  in dense quark matter}, Rev. Mod. Phys. 80 (2008) 1455--1515.
\newblock \href {http://arxiv.org/abs/0709.4635} {\path{arXiv:0709.4635}},
  \href {https://doi.org/10.1103/RevModPhys.80.1455}
  {\path{doi:10.1103/RevModPhys.80.1455}}.

\bibitem{Ohnishi:2015fhj}
A.~Ohnishi, {Approaches to QCD phase diagram; effective models, strong-coupling
  lattice QCD, and compact stars}, J. Phys. Conf. Ser. 668~(1) (2016) 012004.
\newblock \href {http://arxiv.org/abs/1512.08468} {\path{arXiv:1512.08468}},
  \href {https://doi.org/10.1088/1742-6596/668/1/012004}
  {\path{doi:10.1088/1742-6596/668/1/012004}}.

\bibitem{Kitazawa:2001ft}
M.~Kitazawa, T.~Koide, T.~Kunihiro, Y.~Nemoto, {Precursor of color
  superconductivity in hot quark matter}, Phys. Rev. D 65 (2002) 091504.
\newblock \href {http://arxiv.org/abs/nucl-th/0111022}
  {\path{arXiv:nucl-th/0111022}}, \href
  {https://doi.org/10.1103/PhysRevD.65.091504}
  {\path{doi:10.1103/PhysRevD.65.091504}}.

\bibitem{Kitazawa:2003cs}
M.~Kitazawa, T.~Koide, T.~Kunihiro, Y.~Nemoto, {Pseudogap of color
  superconductivity in heated quark matter}, Phys. Rev. D 70 (2004) 056003.
\newblock \href {http://arxiv.org/abs/hep-ph/0309026}
  {\path{arXiv:hep-ph/0309026}}, \href
  {https://doi.org/10.1103/PhysRevD.70.056003}
  {\path{doi:10.1103/PhysRevD.70.056003}}.

\bibitem{Kitazawa:2005vr}
M.~Kitazawa, T.~Koide, T.~Kunihiro, Y.~Nemoto, {Pre-critical phenomena of
  two-flavor color superconductivity in heated quark matter: Diquark-pair
  fluctuations and non-Fermi liquid behavior}, Prog. Theor. Phys. 114 (2005)
  117--155.
\newblock \href {http://arxiv.org/abs/hep-ph/0502035}
  {\path{arXiv:hep-ph/0502035}}, \href {https://doi.org/10.1143/PTP.114.117}
  {\path{doi:10.1143/PTP.114.117}}.

\bibitem{skocpol1975fluctuations}
W.~Skocpol, M.~Tinkham, Fluctuations near superconducting phase transitions,
  Reports on Progress in Physics 38~(9) (1975) 1049.

\bibitem{book_Larkin}
A.~Larkin, A.~Varlamov, Fluctuation phenomena in superconductors, in:
  Superconductivity, Springer, 2008, pp. 369--458.

\bibitem{Abuki:2001be}
H.~Abuki, T.~Hatsuda, K.~Itakura, {Structural change of Cooper pairs and
  momentum dependent gap in color superconductivity}, Phys. Rev. D 65 (2002)
  074014.
\newblock \href {http://arxiv.org/abs/hep-ph/0109013}
  {\path{arXiv:hep-ph/0109013}}, \href
  {https://doi.org/10.1103/PhysRevD.65.074014}
  {\path{doi:10.1103/PhysRevD.65.074014}}.

\bibitem{Voskresensky:2003wd}
D.~N. Voskresensky, {Fluctuations of the color superconducting order parameter
  in heated and dense quark matter} (6 2003).
\newblock \href {http://arxiv.org/abs/nucl-th/0306077}
  {\path{arXiv:nucl-th/0306077}}.

\bibitem{Kunihiro:2007bx}
T.~Kunihiro, M.~Kitazawa, Y.~Nemoto, {How do diquark fluctuations and chiral
  soft modes affect di-lepton production in the deconfined phase?}, PoS CPOD07
  (2007) 041.
\newblock \href {http://arxiv.org/abs/0711.4429} {\path{arXiv:0711.4429}},
  \href {https://doi.org/10.22323/1.047.0041} {\path{doi:10.22323/1.047.0041}}.

\bibitem{Kerbikov:2014ofa}
B.~O. Kerbikov, M.~A. Andreichikov, {Electrical Conductivity of Dense Quark
  Matter with Fluctuations and Magnetic Field Included}, Phys. Rev. D 91~(7)
  (2015) 074010.
\newblock \href {http://arxiv.org/abs/1410.3413} {\path{arXiv:1410.3413}},
  \href {https://doi.org/10.1103/PhysRevD.91.074010}
  {\path{doi:10.1103/PhysRevD.91.074010}}.

\bibitem{Kerbikov:2020lqm}
B.~O. Kerbikov, {Precritical soft photon emission from quark matter}, Phys.
  Rev. D 102~(9) (2020) 096022.
\newblock \href {http://arxiv.org/abs/2001.11766} {\path{arXiv:2001.11766}},
  \href {https://doi.org/10.1103/PhysRevD.102.096022}
  {\path{doi:10.1103/PhysRevD.102.096022}}.

\bibitem{Jaikumar:2001jq}
P.~Jaikumar, R.~Rapp, I.~Zahed, {Photon and dilepton emission rates from high
  density quark matter}, Phys. Rev. C 65 (2002) 055205.
\newblock \href {http://arxiv.org/abs/hep-ph/0112308}
  {\path{arXiv:hep-ph/0112308}}, \href
  {https://doi.org/10.1103/PhysRevC.65.055205}
  {\path{doi:10.1103/PhysRevC.65.055205}}.

\bibitem{McLerran:1984ay}
L.~D. McLerran, T.~Toimela, {Photon and Dilepton Emission from the Quark -
  Gluon Plasma: Some General Considerations}, Phys. Rev. D 31 (1985) 545.
\newblock \href {https://doi.org/10.1103/PhysRevD.31.545}
  {\path{doi:10.1103/PhysRevD.31.545}}.

\bibitem{Weldon:1990iw}
H.~A. Weldon, {Reformulation of finite temperature dilepton production}, Phys.
  Rev. D 42 (1990) 2384--2387.
\newblock \href {https://doi.org/10.1103/PhysRevD.42.2384}
  {\path{doi:10.1103/PhysRevD.42.2384}}.

\bibitem{Kapusta:1991qp}
J.~I. Kapusta, P.~Lichard, D.~Seibert, {High-energy photons from quark - gluon
  plasma versus hot hadronic gas}, Phys. Rev. D 44 (1991) 2774--2788, [Erratum:
  Phys.Rev.D 47, 4171 (1993)].
\newblock \href {https://doi.org/10.1103/PhysRevD.47.4171}
  {\path{doi:10.1103/PhysRevD.47.4171}}.

\bibitem{Hatsuda:1994pi}
T.~Hatsuda, T.~Kunihiro, {QCD phenomenology based on a chiral effective
  Lagrangian}, Phys. Rept. 247 (1994) 221--367.
\newblock \href {http://arxiv.org/abs/hep-ph/9401310}
  {\path{arXiv:hep-ph/9401310}}, \href
  {https://doi.org/10.1016/0370-1573(94)90022-1}
  {\path{doi:10.1016/0370-1573(94)90022-1}}.

\bibitem{Buballa:2003qv}
M.~Buballa, {NJL model analysis of quark matter at large density}, Phys. Rept.
  407 (2005) 205--376.
\newblock \href {http://arxiv.org/abs/hep-ph/0402234}
  {\path{arXiv:hep-ph/0402234}}, \href
  {https://doi.org/10.1016/j.physrep.2004.11.004}
  {\path{doi:10.1016/j.physrep.2004.11.004}}.

\bibitem{Matsuura:2003md}
T.~Matsuura, K.~Iida, T.~Hatsuda, G.~Baym, {Thermal fluctuations of gauge
  fields and first order phase transitions in color superconductivity}, Phys.
  Rev. D 69 (2004) 074012.
\newblock \href {http://arxiv.org/abs/hep-ph/0312042}
  {\path{arXiv:hep-ph/0312042}}, \href
  {https://doi.org/10.1103/PhysRevD.69.074012}
  {\path{doi:10.1103/PhysRevD.69.074012}}.

\bibitem{Giannakis:2004xt}
I.~Giannakis, D.-f. Hou, H.-c. Ren, D.~H. Rischke, {Gauge field fluctuations
  and first-order phase transition in color superconductivity}, Phys. Rev.
  Lett. 93 (2004) 232301.
\newblock \href {http://arxiv.org/abs/hep-ph/0406031}
  {\path{arXiv:hep-ph/0406031}}, \href
  {https://doi.org/10.1103/PhysRevLett.93.232301}
  {\path{doi:10.1103/PhysRevLett.93.232301}}.

\bibitem{Noronha:2006cz}
J.~L. Noronha, H.-c. Ren, I.~Giannakis, D.~Hou, D.~H. Rischke, {Absence of the
  London limit for the first-order phase transition to a color superconductor},
  Phys. Rev. D 73 (2006) 094009.
\newblock \href {http://arxiv.org/abs/hep-ph/0602218}
  {\path{arXiv:hep-ph/0602218}}, \href
  {https://doi.org/10.1103/PhysRevD.73.094009}
  {\path{doi:10.1103/PhysRevD.73.094009}}.

\bibitem{Fejos:2019oxz}
G.~Fej\H{o}s, N.~Yamamoto, {Functional renormalization group approach to color
  superconducting phase transition}, JHEP 12 (2019) 069.
\newblock \href {http://arxiv.org/abs/1908.03535} {\path{arXiv:1908.03535}},
  \href {https://doi.org/10.1007/JHEP12(2019)069}
  {\path{doi:10.1007/JHEP12(2019)069}}.

\bibitem{thouless1960perturbation}
D.~J. Thouless, Perturbation theory in statistical mechanics and the theory of
  superconductivity, Annals of Physics 10~(4) (1960) 553--588.

\bibitem{Cyrot:1973}
M.~Cyrot, Ginzburg-landau theory for superconductors, Reports on Progress in
  Physics 36~(2) (1973) 103.

\bibitem{AL:1968}
L.~Aslamazov, A.~Larkin, Soviet solid state 10, 875 (1968), Phys. Lett. A 26
  (1968) 238.

\bibitem{Maki:1968}
K.~Maki, Critical fluctuation of the order parameter in a superconductor. {I},
  Progress of Theoretical Physics 40~(2) (1968) 193--200.

\bibitem{Thompson:1968}
R.~S. Thompson, Microwave, flux flow, and fluctuation resistance of dirty
  type-{II} superconductors, Physical Review B 1~(1) (1970) 327.

\bibitem{book_Kapusta}
J.~I. Kapusta, C.~Gale, {Finite-temperature field theory: Principles and
  applications}, Cambridge Monographs on Mathematical Physics, Cambridge
  University Press, 2011.
\newblock \href {https://doi.org/10.1017/CBO9780511535130}
  {\path{doi:10.1017/CBO9780511535130}}.

\bibitem{book_LeBellac}
M.~L. Bellac, {Thermal Field Theory}, Cambridge Monographs on Mathematical
  Physics, Cambridge University Press, 2011.
\newblock \href {https://doi.org/10.1017/CBO9780511721700}
  {\path{doi:10.1017/CBO9780511721700}}.

\bibitem{Nara:2021fuu}
Y.~Nara, A.~Ohnishi, {JAM mean-field update: mean-field effects on collective
  flow in high-energy heavy-ion collisions at $\sqrt{s_{NN}}=2-20$ GeV
  energies} (9 2021).
\newblock \href {http://arxiv.org/abs/2109.07594} {\path{arXiv:2109.07594}}.

\bibitem{Rapp:2009yu}
R.~Rapp, J.~Wambach, H.~van Hees, {The Chiral Restoration Transition of QCD and
  Low Mass Dileptons}, Landolt-Bornstein 23 (2010) 134.
\newblock \href {http://arxiv.org/abs/0901.3289} {\path{arXiv:0901.3289}},
  \href {https://doi.org/10.1007/978-3-642-01539-7_6}
  {\path{doi:10.1007/978-3-642-01539-7_6}}.

\bibitem{Laine:2013vma}
M.~Laine, {NLO thermal dilepton rate at non-zero momentum}, JHEP 11 (2013) 120.
\newblock \href {http://arxiv.org/abs/1310.0164} {\path{arXiv:1310.0164}},
  \href {https://doi.org/10.1007/JHEP11(2013)120}
  {\path{doi:10.1007/JHEP11(2013)120}}.

\bibitem{Ghiglieri:2014kma}
J.~Ghiglieri, G.~D. Moore, {Low Mass Thermal Dilepton Production at NLO in a
  Weakly Coupled Quark-Gluon Plasma}, JHEP 12 (2014) 029.
\newblock \href {http://arxiv.org/abs/1410.4203} {\path{arXiv:1410.4203}},
  \href {https://doi.org/10.1007/JHEP12(2014)029}
  {\path{doi:10.1007/JHEP12(2014)029}}.

\bibitem{Adamova:2019vkf}
D.~Adamov\'a, et~al., {A next-generation LHC heavy-ion experiment} (1 2019).
\newblock \href {http://arxiv.org/abs/1902.01211} {\path{arXiv:1902.01211}}.

\bibitem{Barabanov:2020jvn}
M.~Y. Barabanov, et~al., {Diquark correlations in hadron physics: Origin,
  impact and evidence}, Prog. Part. Nucl. Phys. 116 (2021) 103835.
\newblock \href {http://arxiv.org/abs/2008.07630} {\path{arXiv:2008.07630}},
  \href {https://doi.org/10.1016/j.ppnp.2020.103835}
  {\path{doi:10.1016/j.ppnp.2020.103835}}.

\end{thebibliography}

\end{document}